\documentclass[reviewcopy]{elsarticle}
\usepackage{array}
\usepackage[reviewcopy]{adndt}
\usepackage{longtable}
\usepackage{mathrsfs}
\usepackage[table,xcdraw]{xcolor}
\usepackage[figuresright]{rotating}
\usepackage{upgreek}
\usepackage{mathpazo}
\usepackage{booktabs}
\usepackage{wrapfig}
\usepackage{lscape}
\usepackage{longtable}
\usepackage{amssymb}
\usepackage{float}

\usepackage{amsmath}

\usepackage{amssymb,bm}

\biboptions{square,sort&compress}
\bibpunct[]{[}{]}{,}{n}{}{;}
\usepackage[capitalize]{cleveref}

\begin{document}
	
	\begin{frontmatter}
		
		\journal{Atomic Data and Nuclear Data Tables}
		
		
		\title{Magnetic and antimagnetic rotational bands data tables}
		
		\author[author1]{J. X. Teng}

		\author[author1,author2,author3]{K. Y. Ma\corref{cor1}}
		\ead{mky@jlu.edu.cn}

		\cortext[cor1]{Corresponding author.}

		\address[author1]{College of Physics, Jilin University, Changchun 130012, China}
		\address[author2]{Yibin Research Institute, Jilin University, Sichuan 644000, China}  \address[author3]{Chongqing Research Institute, Jilin University, Chongqing, 400000, China}

		\date{19.12.2022} 
		
		\begin{abstract}  
			
			The experimental results of 252 magnetic rotational bands reported in 123 nuclei and 38 antimagnetic rotational bands reported in 27 nuclei are collected and listed in the present work, including energy, spin, parity, magnetic dipole reduced transition probability $B(M1)$, electric quadrupole reduced transition probability $B(E2)$, $B(M1)/B(E2)$ ratio, and the ratio of the dynamic moment of inertia to the electric quadrupole reduced transition probability $\mathcal{J}$$^{(2)}$/$B(E2)$. Following the presentation of the kinematic moment of inertia $\mathcal{J}$$^{(1)}$, dynamic moment of inertia $\mathcal{J}$$^{(2)}$, and $I$ versus rotational frequency $\omega$, as well as energy staggering parameter $S(I)$, $B(M1)$, $B(E2)$, $B(M1)/B(E2)$, and $\mathcal{J}$$^{(2)}$/$B(E2)$ versus $I$ in $A$ $\sim$ 60, 80, 110, 140, and 190 mass regions, a brief discussion is provided. Based on the systematic studies, some nuclei are predicted to be candidates for magnetic or antimagnetic rotation.

			
		\end{abstract}
		
	\end{frontmatter}
	
	
	
	
	\newpage
	
	\tableofcontents
	\listofDtables
	\listofDfigures
	\vskip5pc

	\clearpage
	\section{Introduction}
	Magnetic rotation (MR) and antimagnetic rotation (AMR) are novel kinds of nuclear rotations occurring in weakly deformed or nearly spherical nuclei, and have attracted enormous attention both experimentally and theoretically over the past decades. For MR band \cite{articleF}, it is characterized by a $\Delta$$I$=1 rotational structure with strongly enhanced $M1$ and weak $E2$ transitions \cite{RN93,RN429,RN104H}. The explanation of such bands was given in terms of the shears mechanism predicted by Frauendorf first \cite{RN425}. In this mechanism, the angular momentum vectors of the high-$j$ proton and neutron angular momenta are nearly perpendicular to each other at the bandhead \cite{RN297}. An increase in the energy and total angular momentum is generated by the alignment of the proton and neutron angular momenta in the high-$j$ orbitals. This process can be regarded as the closing of two blades of a pair of shears. Thus, the MR band is also called the shears band \cite{RN341,RN429}. A major feature of magnetic rotation is found under this interpretation: the component of the magnetic dipole vector $\mu_{\perp}$ is perpendicular to total angular momentum at the bandhead and decreases as the spin vectors align, leading to the decrease of the magnetic dipole reduced transition probability $B(M1)$ as the blades close. 


	
	The general properties of magnetic rotation are summarized as \cite{RN93,RN429,RN104H,RN430,RN125}:

	(1) The states in the MR bands generally follow a rotational-like behavior with the energies following the pattern of $E(I) - E_{0}$ $\sim$ A$(I - I_{0})^{2}$, where $I$ is the spin of the state, $I_{0}$ is the spin of the bandhead and A is constant;
	
	(2) The magnetic dipole reduced transition probability $B(M1)$ value is large ($\sim$ $2-10$ $\mu_{N}^{2}$) and decreases. 
	
	(3) The electric quadrupole $(E2)$ transitions are weak or unobserved, and values of electric quadrupole reduced transition probability $B(E2)$ are very small. The $B(M1)/B(E2)$ ratio values are large ($\gtrsim$ 20 [$\mu_{N}$/(eb)]$^{2}$);
	
	(4) The quadrupole deformation parameter is small and less than 0.15 in most cases;
	
	(5) The ratio of the dynamic moment of inertia $\mathcal{J}^{(2)}$ to the $B(E2)$ values are large (>100 MeV$^{{-}1}$(eb)$^{{-}2}$) compared with the values in well deformed ($\sim$ 10 MeV$^{{-}1}$(eb)$^{{-}2}$) or superdeformed bands ($\sim$ 5 MeV$^{{-}1}$(eb)$^{{-}2}$).
	
	MR bands have been investigated theoretically using various models including shell model \cite{FRAUENDORF199641} and the many-particles-plus-rotor models \cite{PhysRevC.74.044310} and the titled axis cranking (TAC) models \cite{RN425} have been proved to be the powerful tools. The validity of TAC approximation was discussed and tested in comparison with particle-rotor model \cite{RN422}. The Skyrme-Hartree-Fock Method with the TAC has been used to investigate MR band in $^{142}$Gd \cite{etde_20302535}. Because of the high numerical complexity of the TAC model, most of the applications are based on simple phenomenological Hamiltonians, such as the pairing-plus-quadrupole-quadrupole tilted-axis cranking (PQTAC) model \cite{PhysRevC.75.044309}. The covariant density functional theory (CDFT) has received a lot of attention during the recent years. In 2000, the three-dimensional cranking CDFT was established by Madokoro et al \cite{PhysRevC.62.061301}. However, it has only been used to investigate magnetic rotation in $^{84}$Rb. In comparison, the two-dimensional cranking CDFT reduces the computing time considerably, and thus can be applied for medium heavy nucleus such as $^{142}$Gd \cite{PhysRevC.78.024313}. Moreover, the TAC model based on the CDFT with the point-coupling interaction was developed by Zhao et al \cite{ZHAO2011181}, which has been successfully applied for MR ranging from light nuclei $^{58}$Fe \cite{RN132}, $^{60}$Ni \cite{ZHAO2011181} to heavy nuclei $^{198,~199}$Pb \cite{PhysRevC.85.024318}.
	
	For AMR band \cite{RN429,RN430}, it is characterized by a $\Delta$$I$=2 rotational structure with a rather regular cascade of weak $E2$ transitions. The energy and higher angular momentum are obtained by the ‘two-shears-like mechanism’, which was first proposed by Frauendorf \cite{RN429}. In this mechanism, the angular momenta of protons (neutrons) are aligned back to back in opposite directions and roughly perpendicular to the orientation of the total angular momentum of neutrons (protons) at the bandhead. Along this band, energy and angular momentum are increased by simultaneous closing of the two blades of protons and neutrons toward the the neutron (proton) angular momentum vector \cite{RN430,RN426}. Under this interpretation, the transverse magnetic moment is zero since $\mu_{\perp}$ of the two subsystems are antialigned and cancel each other.

	The experimental indicators for AMR bands can be summarized as:

	(1)  A $\Delta$$I$=2 sequence with $E2$ transitions only, as the cancellation of the magnetic moments leads to the absence of the $M1$ transition;
	
	(2) The structures are weakly deformed leading to small $B(E2)$ values, which decrease with the increasing spin;
	
	(3) The ratio of the dynamic moment of inertia to electric quadrupole reduced transition probability $\mathcal{J}$$^{(2)}$/$B(E2)$ values are large (> 100 $\hbar^{2}$MeV$^{-1}$(eb)$^{-2}$). 
	
	From the theoretical aspect, antimagnetic rotation has been discussed mainly by the semiclassical particle rotor model \cite{RN93} and TAC model \cite{FRAUENDORF2000115,PhysRevC.78.024313,ZHAO2011181}. Many investigations in the framework of microscopic-macroscopic model \cite{RN386,RN388,RN371}, pairing plus quadrupole model \cite{RN83,RN429}, and the CDFT \cite{RN426,PhysRevC.85.054310,RN106L,RN395,RN115,RN375} with the point coupling effective interaction PC-PK1 \cite{PhysRevC.82.054319} have been carried out based on the TAC model. The cranked shell model (CSM) with pairing correlations treated by a particle-number-conserving 
	(PNC) method \cite{PhysRevC.50.1388,ZENG19831}, is also successfully used to investigate AMR bands.
	
\begin{figure}[H]
		\centering
		\includegraphics[width=15cm]{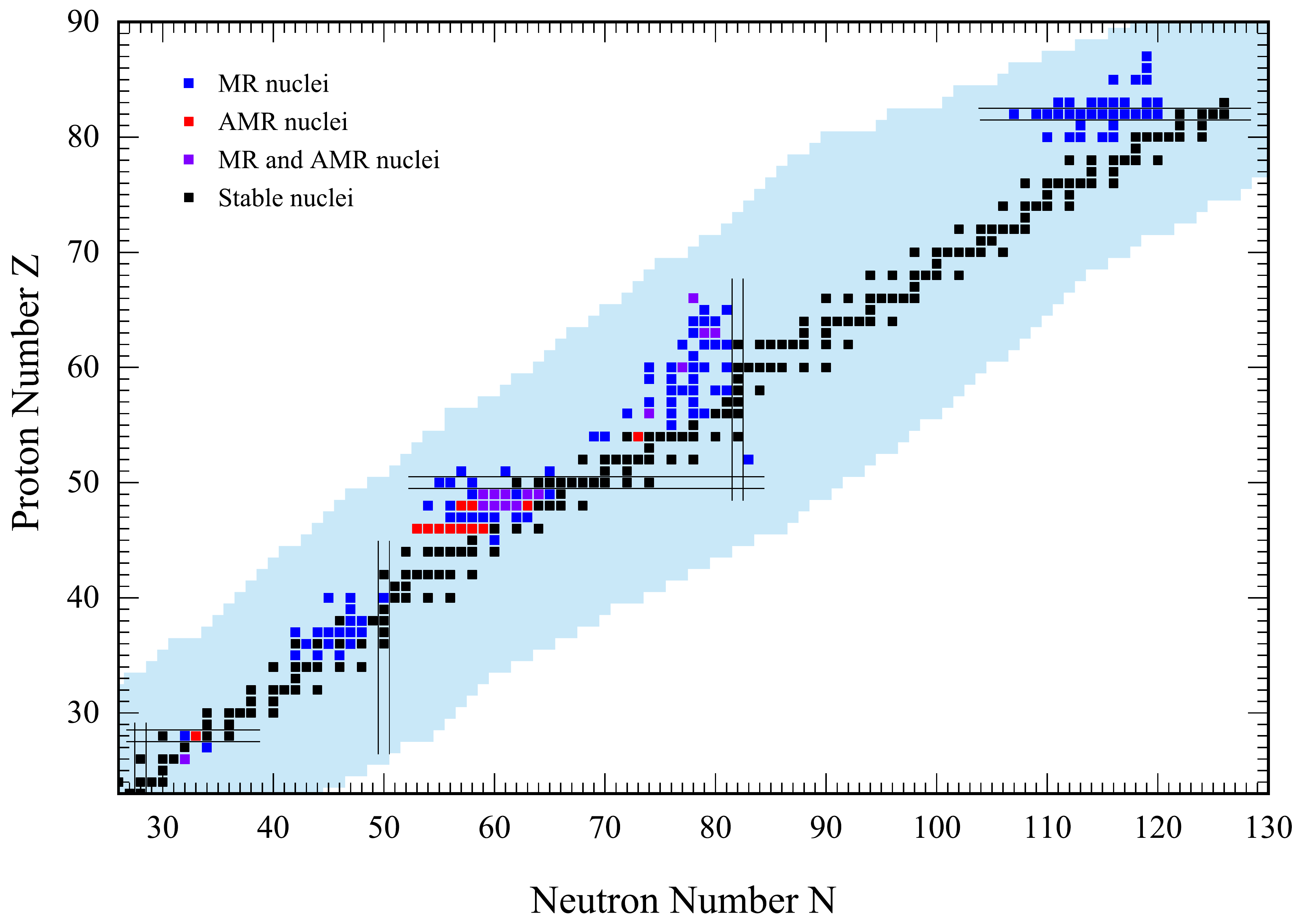}
		\caption{(color online) On the nuclear chart, the nuclides with magnetic rotation and antimagnetic rotation are observed. The blue squares represent the nuclei of magnetic rotation only. The red squares represent the nuclei of antimagnetic rotation only. The purple squares represent the coexistence of both magnetic and antimagnetic rotations. The light blue background represents the currently observable nuclei.}\label{fig.1}
	\end{figure}

	Both magnetic and antimagnetic rotations have been investigated in experimental and theoretical research for many years. Since the clear evidence for shears mechanism has first been provided through lifetime measurements for four $M1$ bands in $^{198,~199}$Pb \cite{RN416}, more and more nuclei are confirmed to have MR or AMR bands. We have expanded the previous compilation of MR bands made by Amita et al. in 2000 \cite{AMITA2000283}. Up to now, 252 magnetic rotational bands in 123 nuclei have been reported in the $A$ $\sim$ 60 \cite{RN132,RN126,RN220,RN131,ZHAO2011181,PhysRevC.107.014307}, 80 \cite{RN167,RN168,RN181,RN176,RN178,RN177,RN180,RN183,SCHWENGNER1990550,RN182,MU2022137006,RN184,RN185,RN187,RN188,RN170,RN190,RN128,RN191,RN192,RN193,RN194,RN144,RN196,RN198,RN199,RN173,PhysRevC.82.014306,RN174,RN369,RN127,RN130,RN129,RUSU20091,RN370,RN223,article85,article851,article853,RN224,Wang_2021,osti_5772960}, 110 \cite{RN68,RN62,RN63,RN64,RN65,RN66,RN67,RN100,RN99,RN98,RN97,PhysRevLett.112.202502,RN101,RN102,Zhang_2011,RN109,RN108,RN113,RN114,RN112,PhysRevC.103.024302,RN60,RN59,Jenkins_2000,Robinson2002HighspinSA,RN103,RN77,RN73,RN74,RN75,RN83,RN82,RN84,RN85,RN115,RN105,RN78,RN111,RN116,RN87,RN120,RN117,RN118,RN119,RN121,RN90,RN91,RN402,RN122,RN123,Li2011LevelSI,RN70,RN69,RN72,RN71,RN79,RN88,RN89,etde_449103,DAR2022122382,RN124}, 140 \cite{RN251,RN252,RN226,RN227,RN229,RN228,RN236,RN230,RN233,RN234,RN232,RN231,RN235,RN239,RN240,RN245,RN248,RN238,RN237,RN206,RN250,PhysRevC.87.034317,RN246,RN249,RN253,RN254,RN255,RN207,RN208,RN243,RN244,RN211,RN219,RN258,RN247,RN256,RN257,RN260,RN259,RN261,RN217,RN263,RN262,RN209,RN266,RN264,RN265,RN203,RN204,RN278,RN202,RN279,RN267,RN419,RN269,RN205,etde_20302535,RN270,RN271,article142,RN272,RN420,RN275,RN276,RN277,RN281,RN140,RN212,RN282,RN283,RN213,RN280}, and 190 \cite{RN287,RN289,RN290,RN415,RN293,RN294,RN306,RN310,RN311,RN216,RN329,RN286,RN285,RN288,RN291,RN295,RN297,RN296,RN298,PhysRevC.79.014315,RN301,RN302,RN303,RN327,etde_184510,RN304,RN307,RN308,RN312,RN313,RN314,RN317,RN316,RN315,RN319,RN333,RN340,RN324,RN325,RN326,RN328,RN322,RN416,RN334,RN335,RN337,RN341,article199,RICHEL1978483,RN339,RN343,RN363,RN362,RN300,RN218,RN320,RN332,RN330,RN417,RN361,RN135,RN364,RN214,RN366} mass regions. In contrast, the experimental observation of AMR bands is scarce and so far only 38 antimagnetic rotational bands are reported in 27 nuclei and mainly distributed in the $A$ $\sim$ 60 \cite{RN9058,RN132,PhysRevC.107.014307}, 110 \cite{Sihotra2022PossibleAR,PhysRevC.83.024313,RN372,RN371,MA2021122319,RN377,RN378,RN373,RN374,RN375,RN376,RN87102,JERRESTAM1996203,RN379,RN380,Pan_2022,PhysRevC.60.024307,RN382,RN381,Majumder_2020,RN92105,article105,RN383,RN426,RN410,RN389,RN386,RN388,RN387,RN103,RN390,RN391,RN397,PhysRevC.49.1885,RN395,JUUTINEN1994727,RN394,RN398,RN400,RN414,RN402,RN73,RN83,RN115,RN78}, and 140 \cite{RN404,RN411,RN405,RN406,RN235,PETRACHE2019241,RN280} mass regions. The nuclei of magnetic and antimagnetic rotations in the nuclear chart are given in \cref{fig.1}.
	
\begin{figure}[H]
		\centering
		\includegraphics[width=15cm]{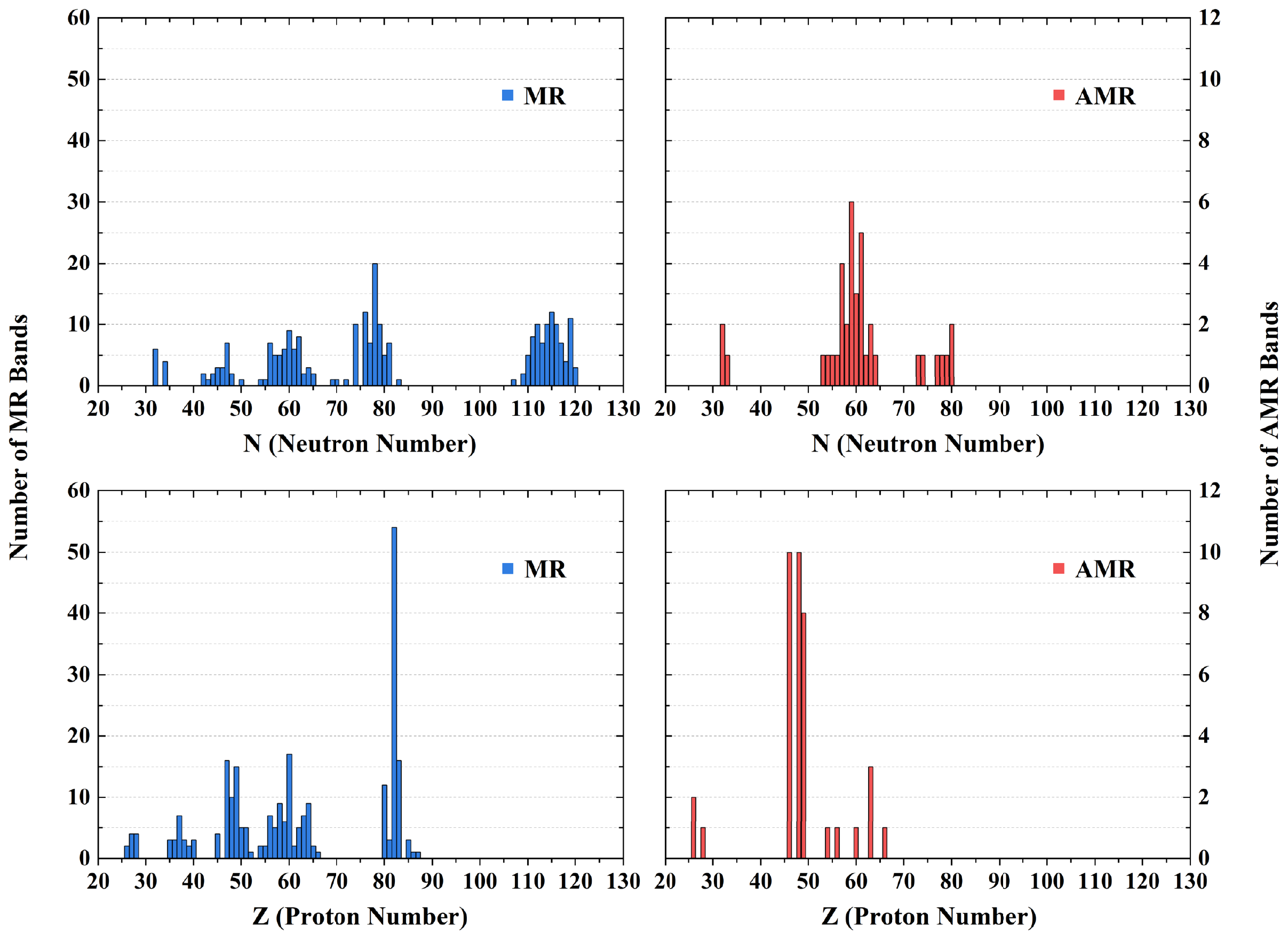}
		\caption{Numbers of MR and AMR bands as a function of neutron and proton number, respectively}\label{NP}
	\end{figure}

	The numbers of observed MR and AMR bands as a function of neutron and proton numbers, respectively, are shown in \cref{NP}. Most MR bands are reported near magic or semi-magic numbers $N$ $=$ 107$-$120, 74$-$83, 54$-$66, 42$-$48, and 32$-$34. Same is also true for protons where MR bands are seen near $Z$ $=$ 80$-$83, 55$-$64, 45$-$51, 35$-$40, and 26$-$28, among which the Pb isotopes ($Z$ $=$ 82) have the largest number of MR bands reported. A majority of AMR bands are reported near numbers $N$ $=$ 77$-$80, 54$-$64, 32, and 33. The largest number of AMR bands are reported in $N$ $=$ 60. In terms of proton number, AMR bands are reported at $Z$ $=$ 26, 28, 46, 48, 49, 54, 56, 60, 63, and 66.
	
	In order to promote the study of magnetic and antimagnetic rotations, the compilation of the data for MR and AMR bands is demanded. In this paper, we mainly focus on the discussion of spin $I$, kinematic moment of inertia $\mathcal{J}$$^{(1)}$, and dynamic moment of inertia $\mathcal{J}$$^{(2)}$ versus rotational frequency $\omega$, as well as energy staggering parameter $S(I)$, the magnetic dipole reduced transition probability $B(M1)$, the electric quadrupole reduced transition probability $B(E2)$ and the $B(M1)/B(E2)$ ratio versus spin $I$ in $A$ $\sim$ 60, 80, 110, 140, and 190 mass regions for MR bands. Moreover, the spin $I$, kinematic moment of inertia $\mathcal{J}$$^{(1)}$, and dynamic moment of inertia $\mathcal{J}$$^{(2)}$ versus rotational frequency $\omega$, as well as the electric quadrupole reduced transition probability $B(E2)$ and the $\mathcal{J}$$^{(2)}$/$B(E2)$ ratio versus spin $I$ for AMR bands in $A$ $\sim$ 60, 110, and 140 mass regions are also discussed. Experimental data on MR and AMR bands are presented in Table A and B, respectively. 

	\section{Systematics of magnetic rotational bands}
	
	\subsection{The spin vs. rotational frequency }
\begin{figure}[H]
		\centering
		\includegraphics[width=18.5cm]{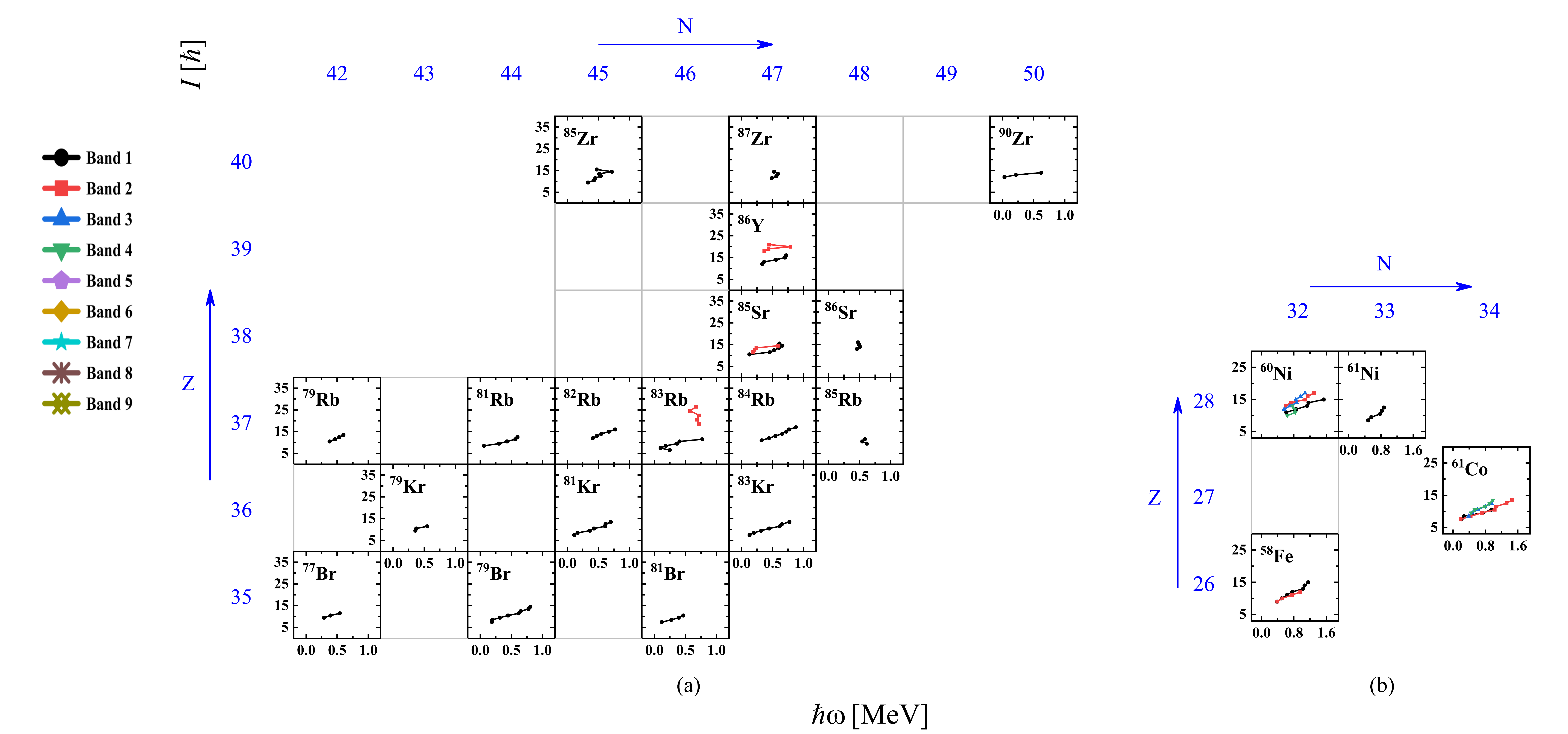}
		\caption{(Color online) Spin versus rotational frequency for magnetic rotational bands in $A$ $\sim$ (a) 80 and (b) 60 mass regions.}
		\label{9}
		
	\end{figure}

\begin{figure}[H]
		\centering
		\includegraphics[width=15.1cm]{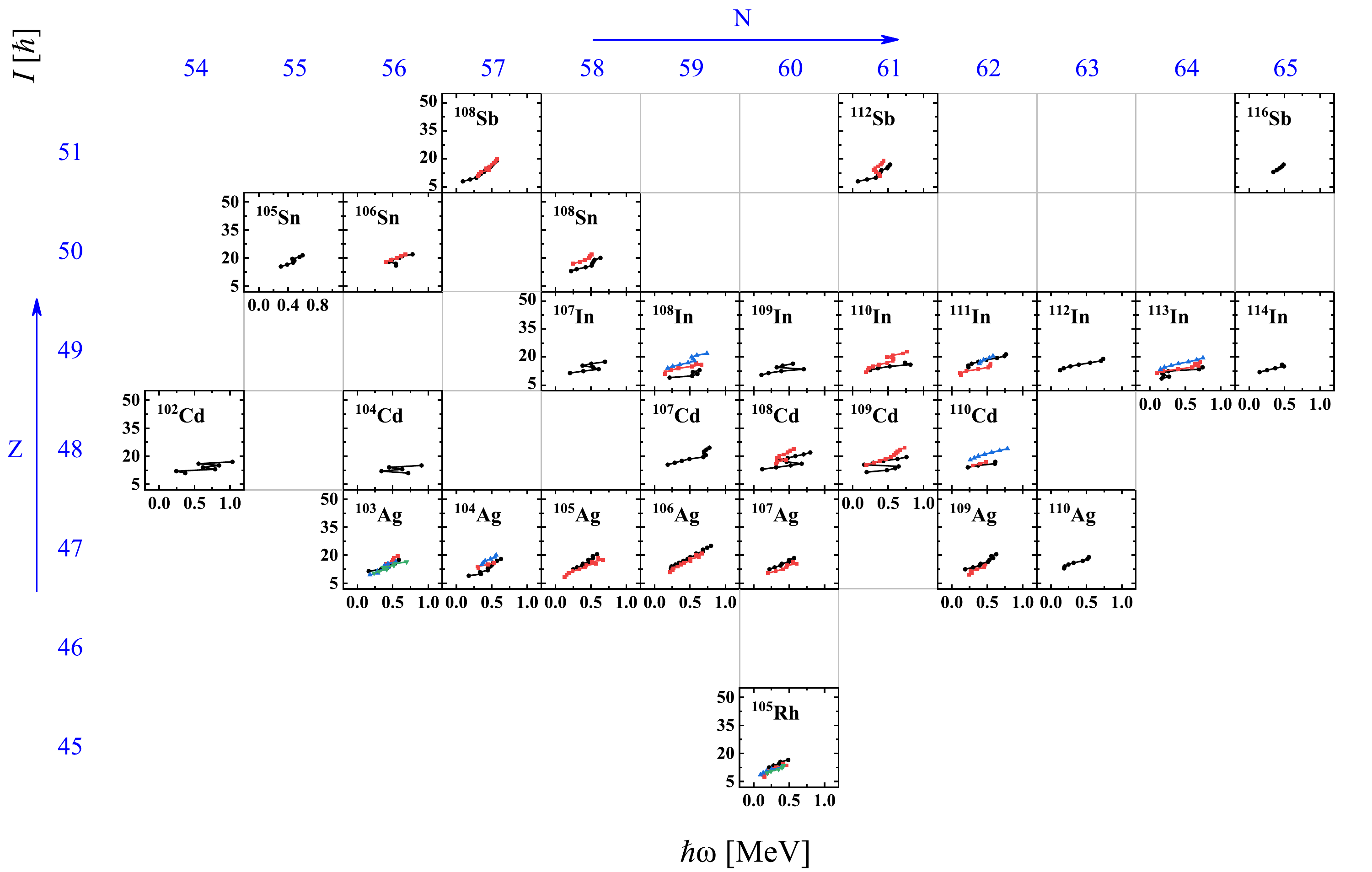}
		\caption{(Color online) Spin versus rotational frequency for magnetic rotational bands in $A$ $\sim$ 110 mass region.}
		\label{10}
	\end{figure}

\begin{figure}[H]
		\centering
		\includegraphics[width=18cm]{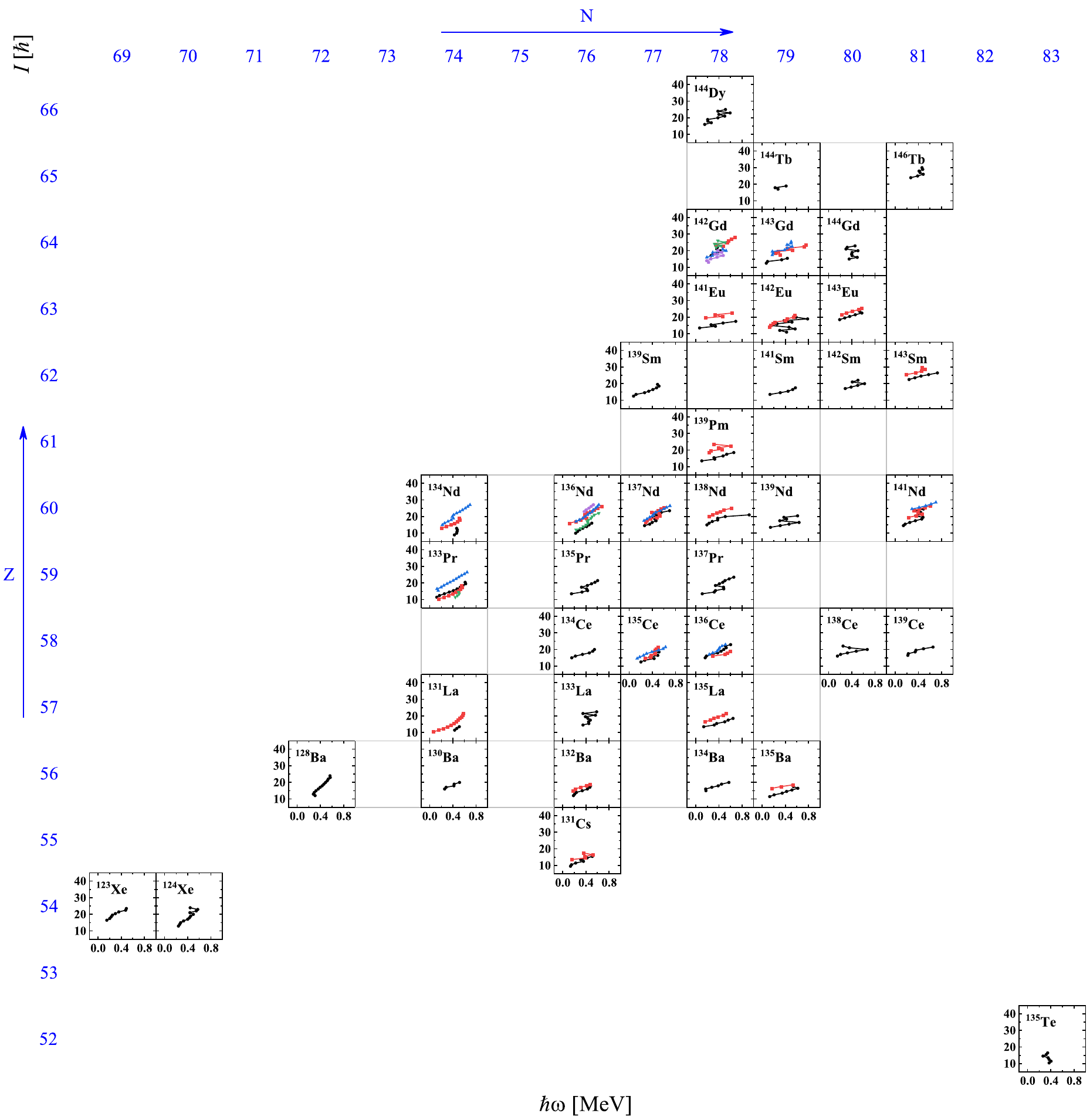}
		\caption{(Color online) Spin versus rotational frequency for magnetic rotational bands in $A$ $\sim$ 140 mass region.}
		\label{11}
	\end{figure}

\begin{figure}[H]
		\centering
		\includegraphics[width=16.5cm]{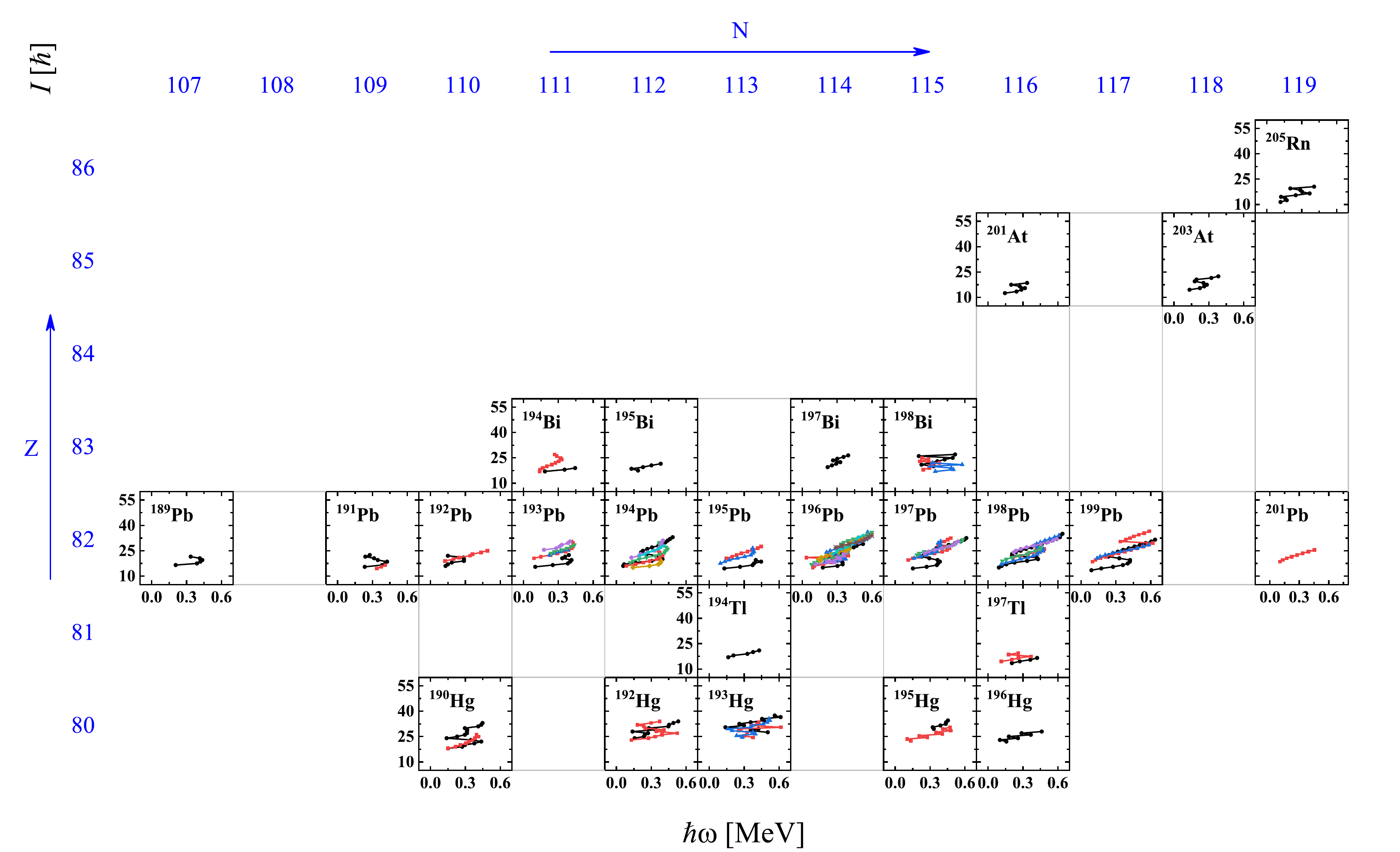}
		\caption{(Color online) Spin versus rotational frequency for magnetic rotational bands in $A$ $\sim$ 190 mass region.}
		\label{12}
	\end{figure}

	\cref{9,10,11,12} show the relations between spins and rotational frequencies for all magnetic rotational bands in $A$ $\sim$ 60, 80, 110, 140, and 190 mass regions, respectively. It is noted that a number of MR bands are observed to display band-crossing phenomenon in $A$ $\sim$ 110, 140, and 190 mass regions from \cref{10,11,12}. However, as shown in \cref{9}, the majority of nuclei do not show the band-crossing phenomenon in $A$ $\sim$ 60 and 80 mass regions. Generally, the spin tends to grow smoothly with increasing rotational frequency before and after the backbending. But, there is an abnormal phenomenon in $^{83}$Rb and $^{135}$Te, where the spin decreases with rotational frequency increasing. Moveover, the magnetic rotation occurs at the lowest value spin 5.5$\hbar$ for $^{83}$Rb, and the highest value spin 37.5$\hbar$ for $^{193}$Hg.
	
	\subsection{The kinematic moment of inertia vs. rotational frequency}
\begin{figure}[H]
		\centering
		\includegraphics[scale=0.26]{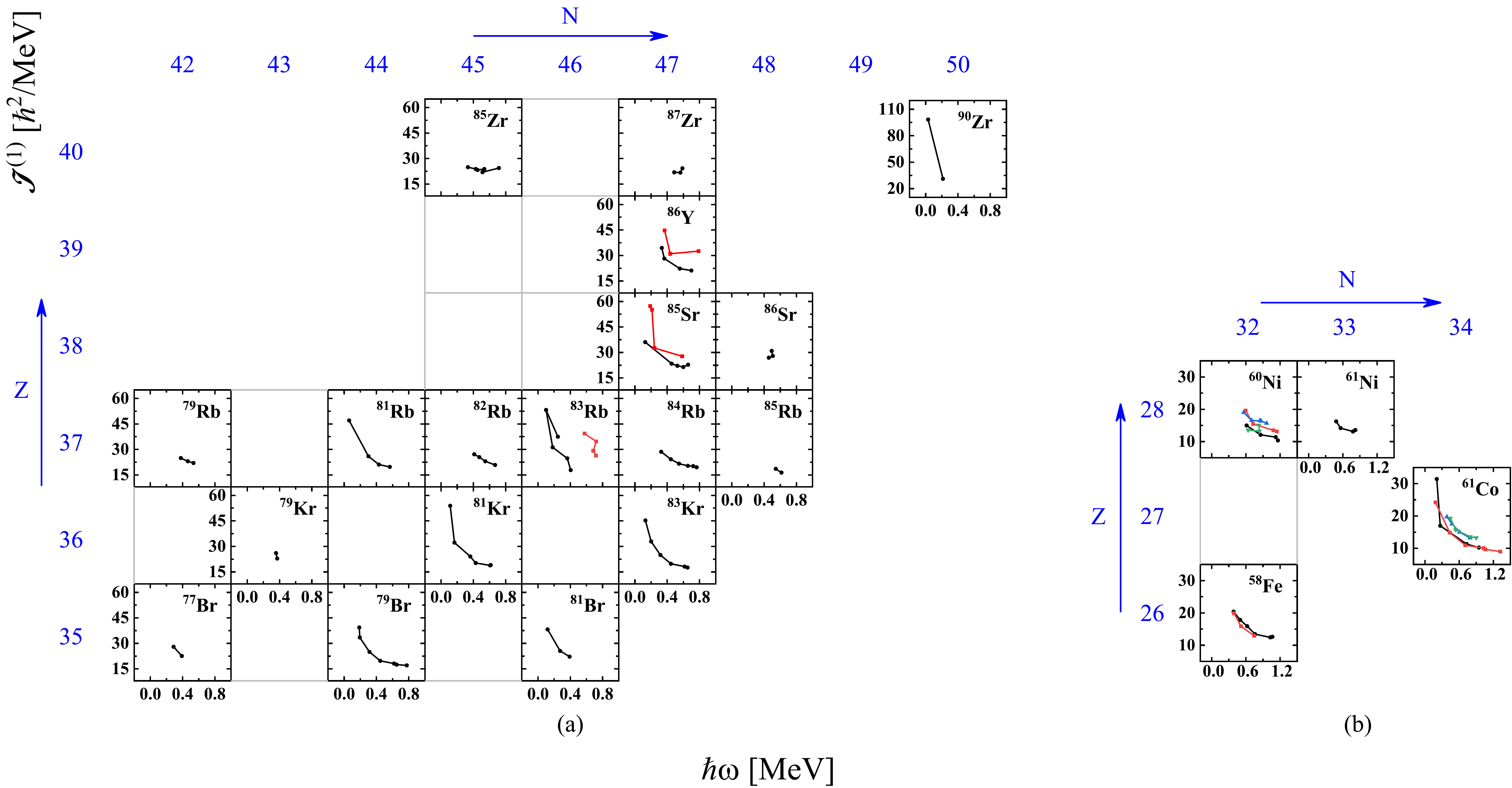}
		\caption{(Color online) Kinematic moment of inertia versus rotational frequency for magnetic rotational bands in $A$ $\sim$ (a) 80 and (b) 60 mass regions.}
		\label{1}
		
	\end{figure}

\begin{figure}[H]
		\centering
		\includegraphics[width=15cm]{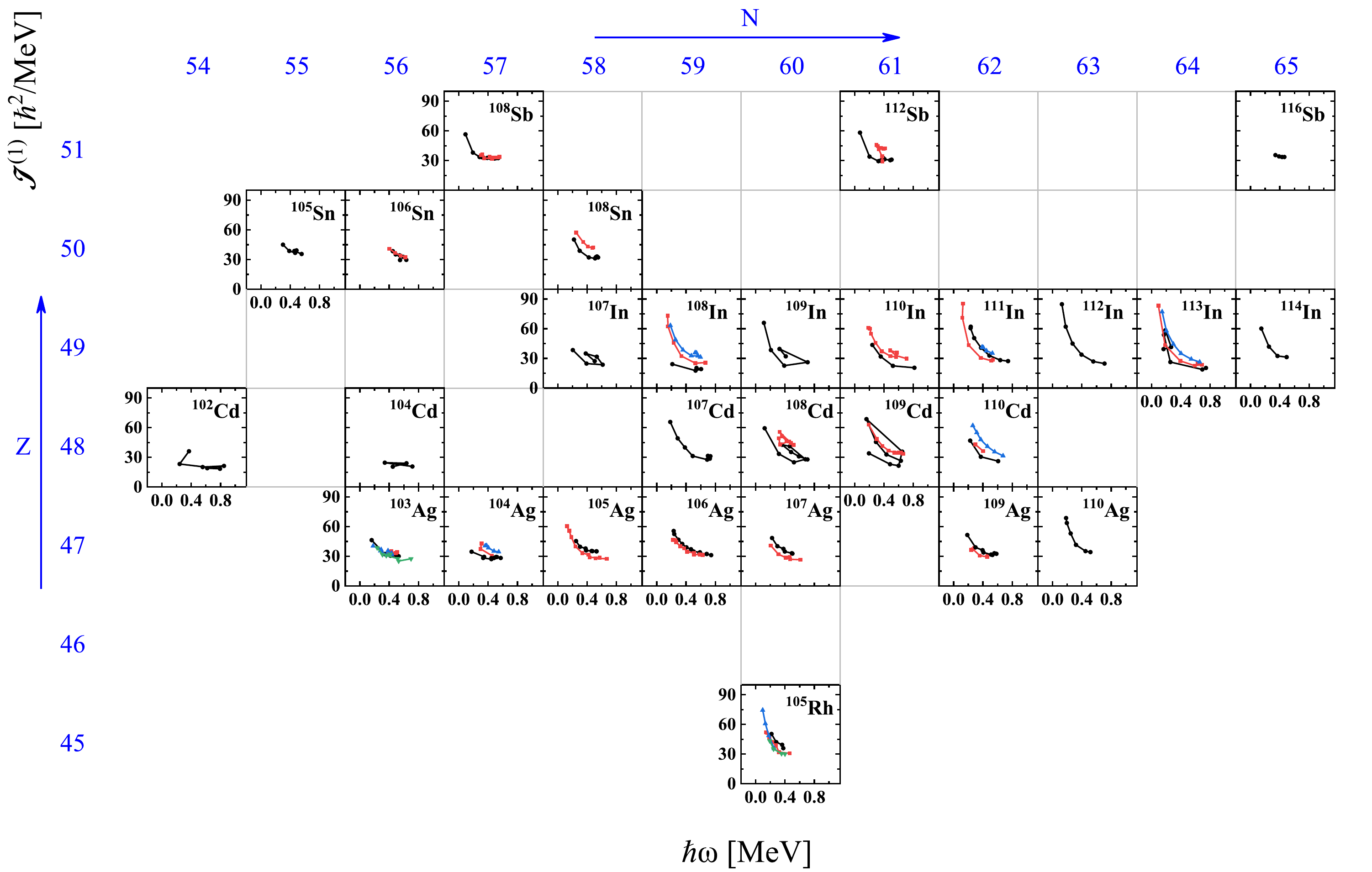}
		\caption{(Color online) Kinematic moment of inertia versus rotational frequency for magnetic rotational bands in $A$ $\sim$ 110 mass region.}
		\label{2}
	\end{figure}

\begin{figure}[H]
		\centering
		\includegraphics[width=18.0cm]{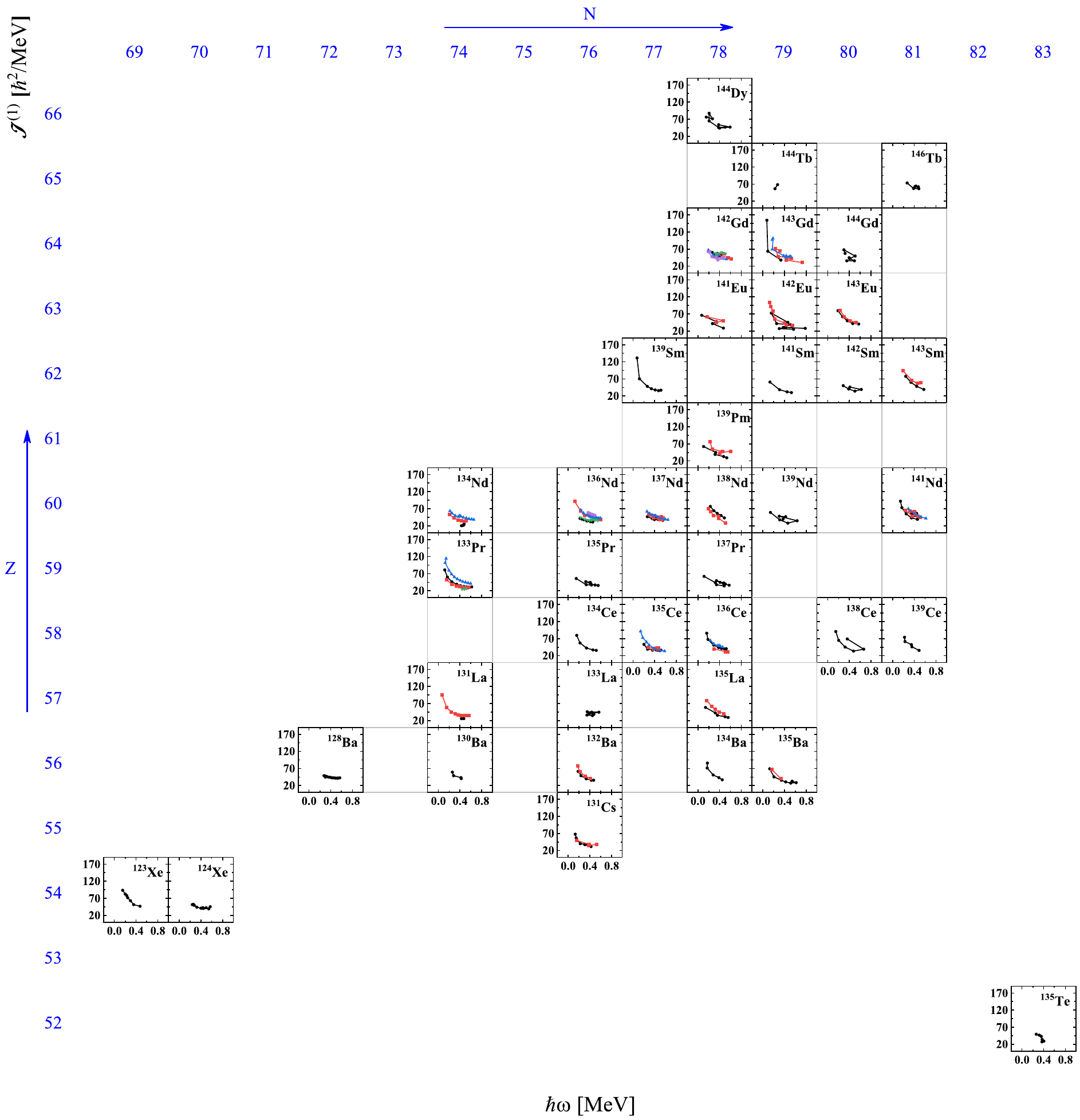}
		\caption{(Color online) Kinematic moment of inertia versus rotational frequency for magnetic rotational bands in $A$ $\sim$ 140 mass region.}
		\label{3}
	\end{figure}

\begin{figure}[H]
		\centering
		\includegraphics[width=16.5cm]{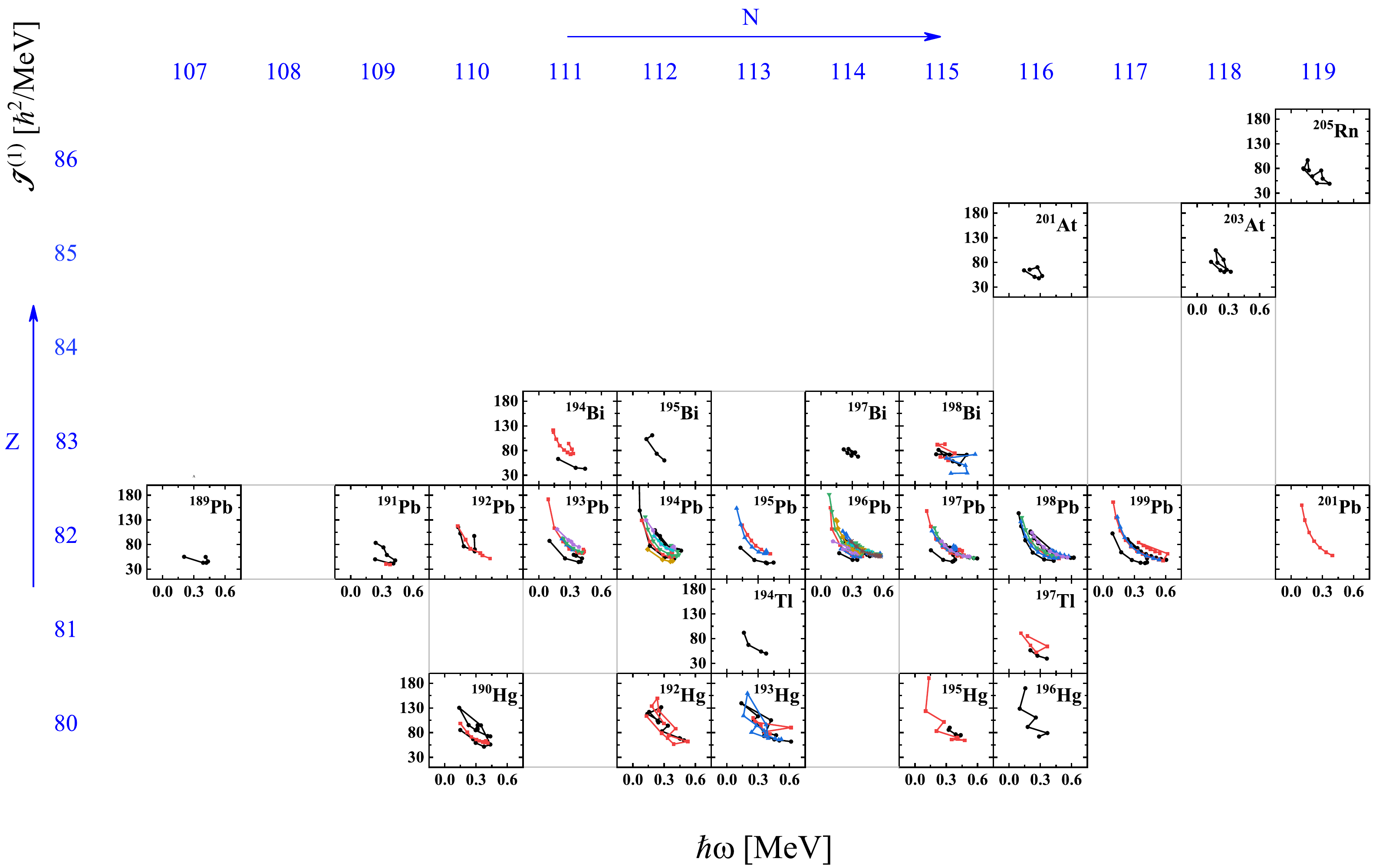}
		\caption{(Color online) Kinematic moment of inertia versus rotational frequency for magnetic rotational bands in $A$ $\sim$ 190 mass region.}
		\label{4}
	\end{figure}

	The kinematic moments of inertia for all magnetic rotational bands in $A$ $\sim$ 60, 80, 110, 140, and 190 mass regions are given in \cref{1,2,3,4}, respectively. The value of $\mathcal{J}$$^{(1)}$ is estimated by using the relation $\mathcal{J}$$^{(1)}$=$I$/$\omega$. It is clear that the kinematic moment of inertia gradually decreases with the rotational frequency increasing before and after the backbending, and the decline speed gradually becomes gentler as the rotational frequency increases. However, one observes an opposite trend for MR bands in $^{86}$Sr, $^{87}$Zr, and $^{144}$Tb where $\mathcal{J}$$^{(1)}$ generally rises with increasing rotational frequency. 
	It can be seen that in the $A$ $\sim$ 60 mass region, $\mathcal{J}$$^{(1)}$ varies from 9 to 31 keV/$\hbar$, in the $A$ $\sim$ 80 mass region, $\mathcal{J}$$^{(1)}$ varies from 9 to 57 keV/$\hbar$ except for $^{90}$Zr, and it varies from 17 to 85 keV/$\hbar$ in the $A$ $\sim$ 110 mass region. Then, in the $A$ $\sim$ 140 mass region, $\mathcal{J}$$^{(1)}$ varies from 20 to 154 keV/$\hbar$, and finally it varies from 39 to 258 keV/$\hbar$ in the $A$ $\sim$ 190 mass region. It is shown that both the upper and lower bounds of the $\mathcal{J}$$^{(1)}$ values gradually increase with the change of the mass region from $A$ $\sim$ 60 to $A$ $\sim$ 190 mass region.
	
	\subsection{The dynamic moment of inertia vs. rotational frequency}
	
\begin{figure}[H]
		\centering
		\includegraphics[scale=0.26]{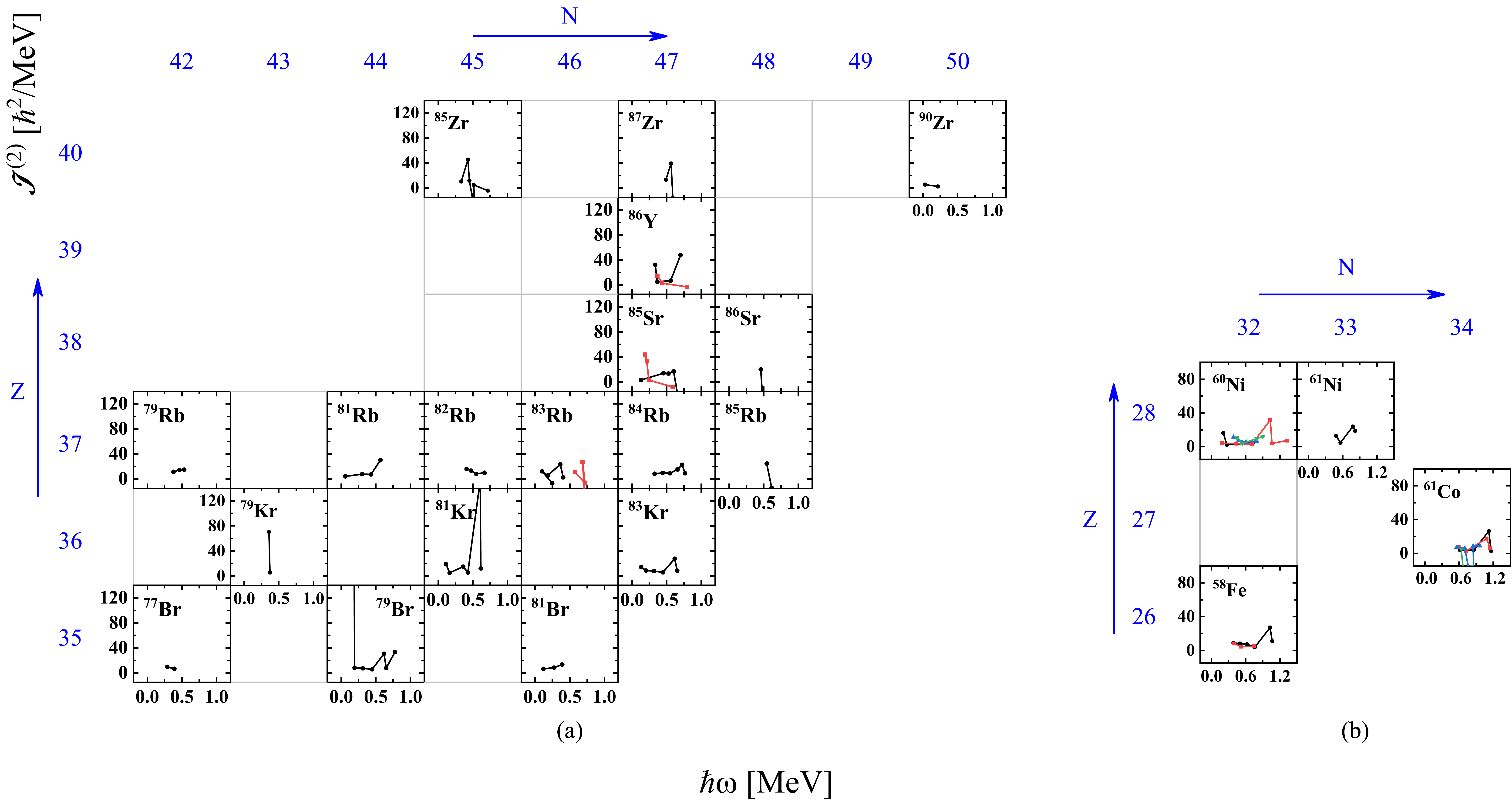}
		\caption{(Color online) Dynamic moment of inertia versus rotational frequency for magnetic rotational bands in $A$ $\sim$ (a) 80 and (b) 60 mass regions.}
		\label{5}
	\end{figure}

\begin{figure}[H]
		\centering
		\includegraphics[width=15cm]{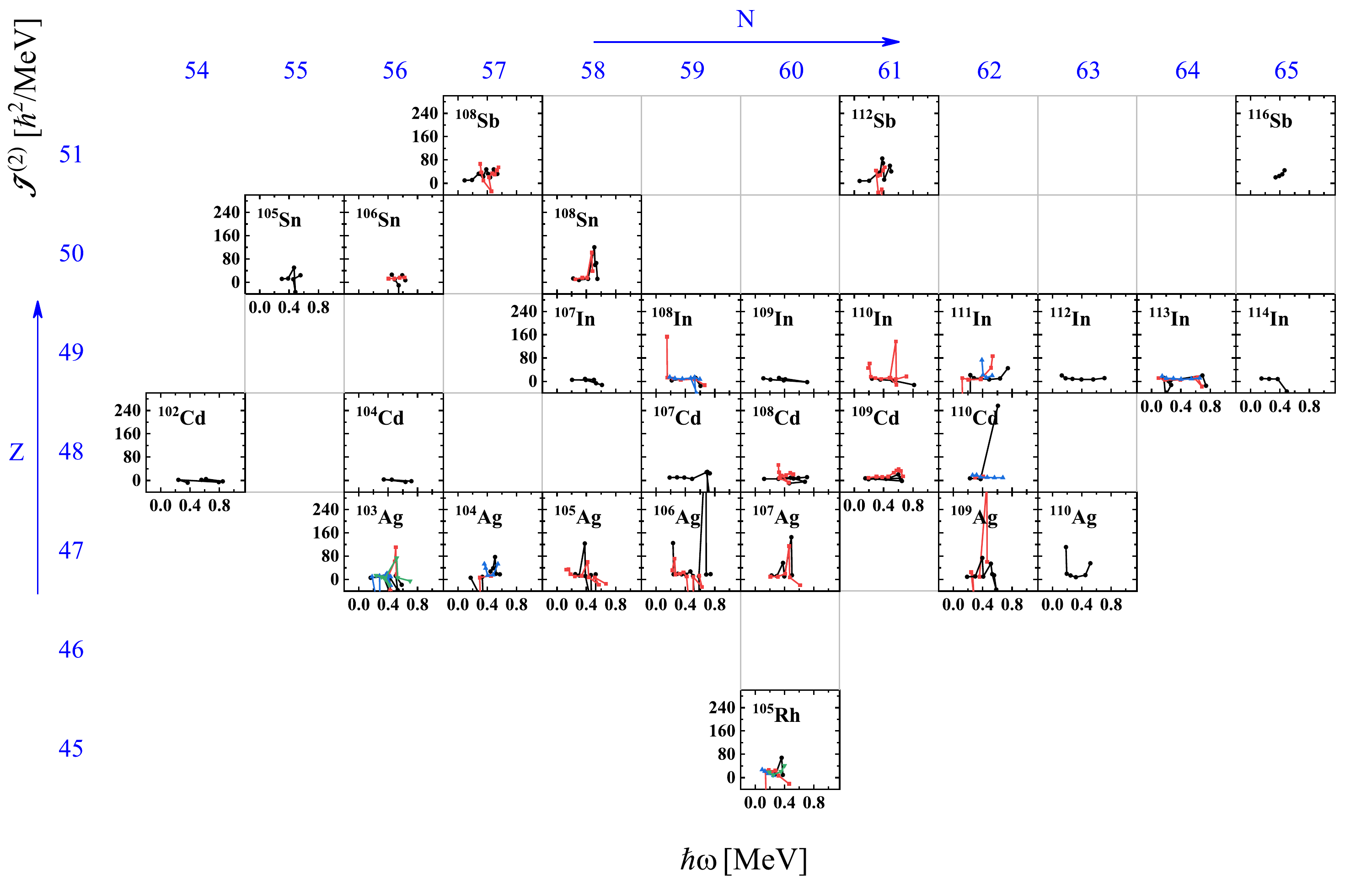}
		\caption{(Color online) Dynamic moment of inertia versus rotational frequency for magnetic rotational bands in $A$ $\sim$ 110 mass region.}
		\label{6}
	\end{figure}

\begin{figure}[H]
		\centering
		\includegraphics[width=18cm]{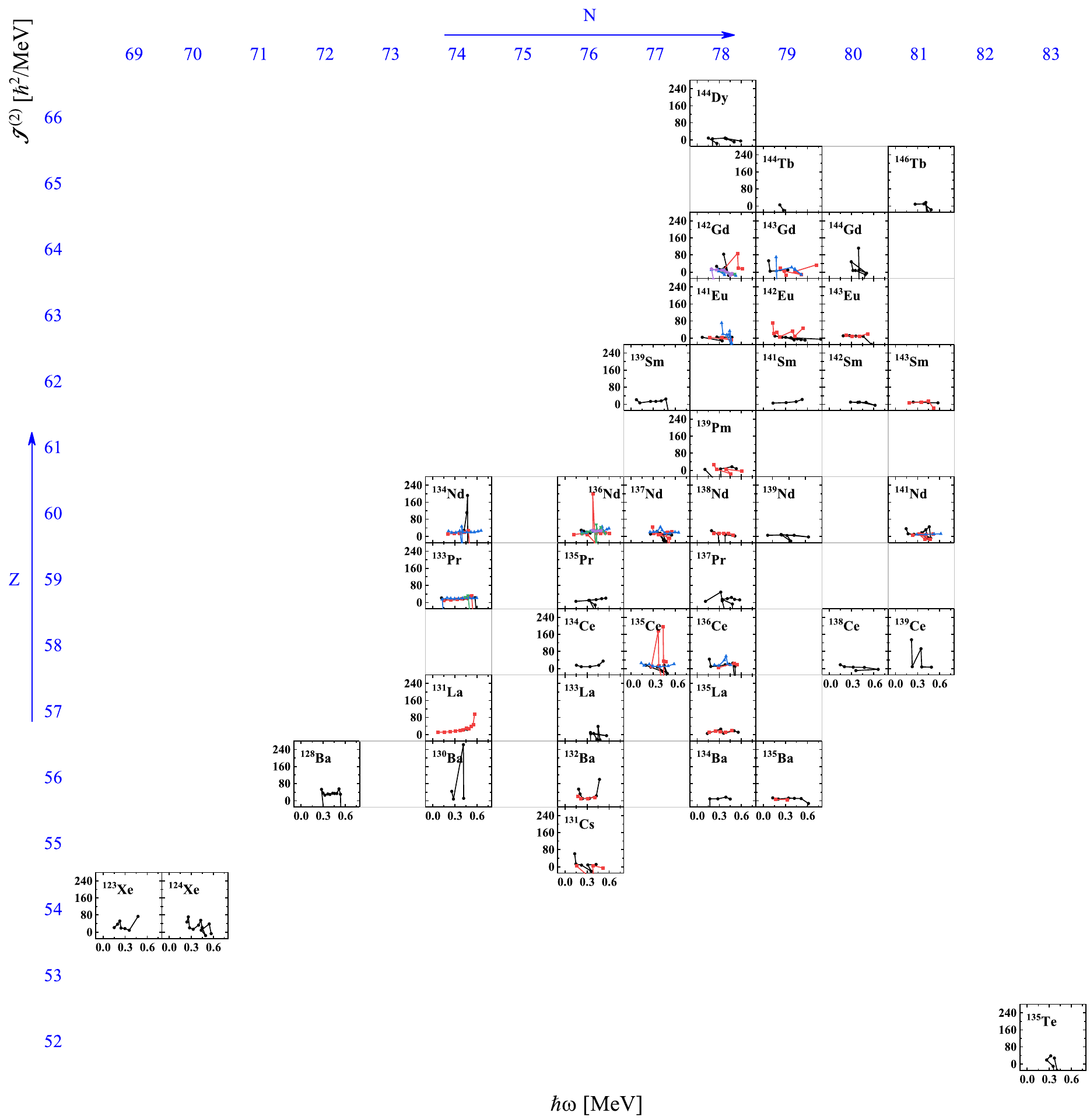}
		\caption{(Color online) Dynamic moment of inertia versus rotational frequency for magnetic rotational bands in $A$ $\sim$ 140 mass region.}
		\label{7}
	\end{figure}

\begin{figure}[H]
		\centering
		\includegraphics[width=16.8cm]{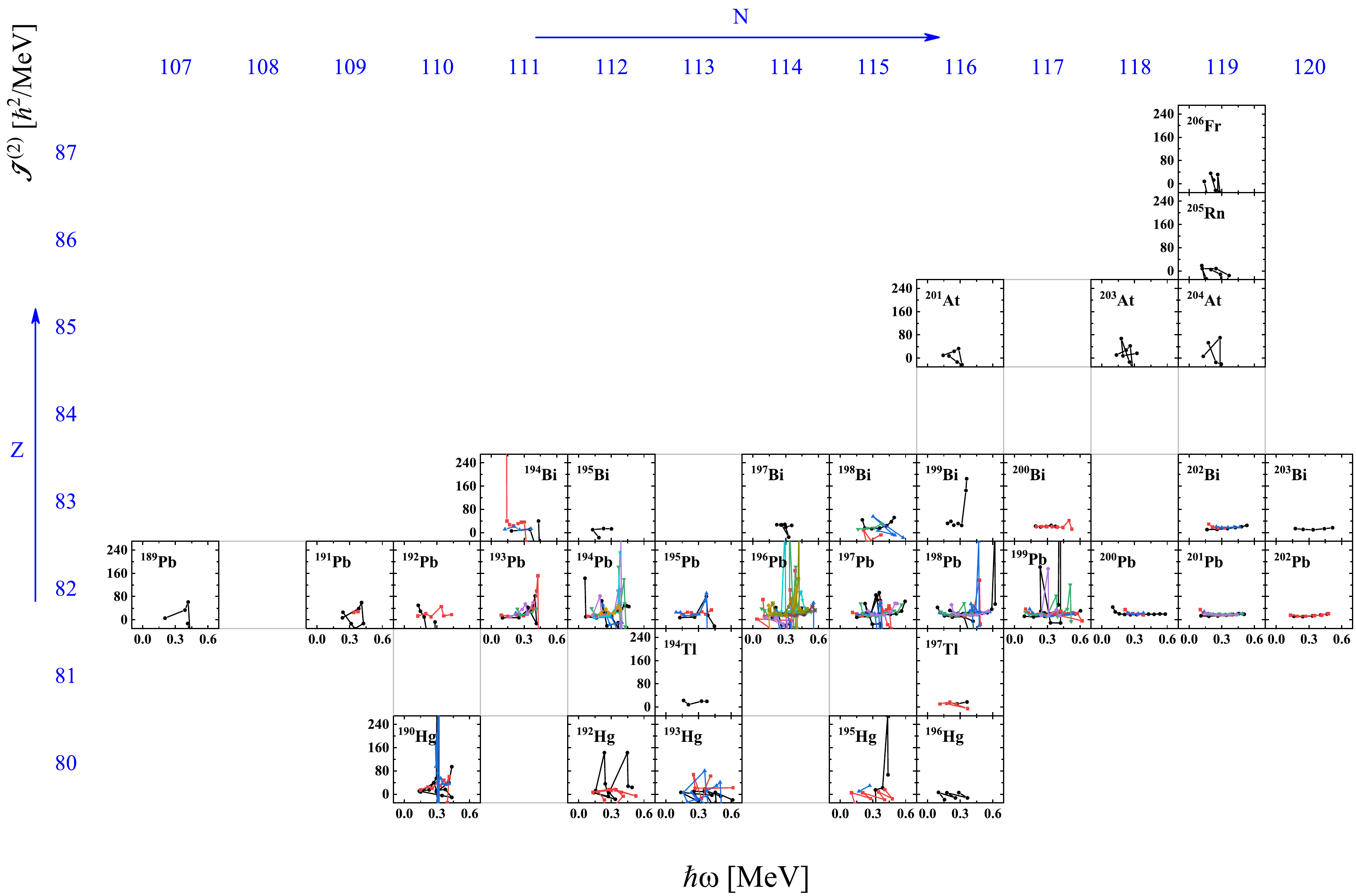}
		\caption{(Color online) Dynamic moment of inertia versus rotational frequency for magnetic rotational bands in $A$ $\sim$ 190 mass region.}
		\label{8}
	\end{figure}

	\cref{5,6,7,8} show the dynamic moments of inertia $\mathcal{J}$$^{(2)}$ for all magnetic rotational bands in A $\sim$ 60, 80, 110, 140, and 190 mass regions, respectively. The value of $\mathcal{J}$$^{(2)}$ is estimated by using the relation $\mathcal{J}$$^{(2)}$=$1/[E_\gamma(I+1 \rightarrow I)-E_\gamma(I \rightarrow I-1$)]. A large fluctuation of $\mathcal{J}$$^{(2)}$ values exists in $A$ $\sim$ 60, 80, 110, 140, and 190 mass regions since the $\mathcal{J}$$^{(2)}$ corresponds to the second derivative of the energy with the spin. Furthermore, it is clear that $\mathcal{J}$$^{(2)}$ values of MR bands are quite small and some of them are roughly constant while others have a large fluctuation  with  rotational frequency increasing.
	
	\subsection{The energy staggering vs. spin}
\begin{figure}[H]
		\centering
		\includegraphics[scale=0.26]{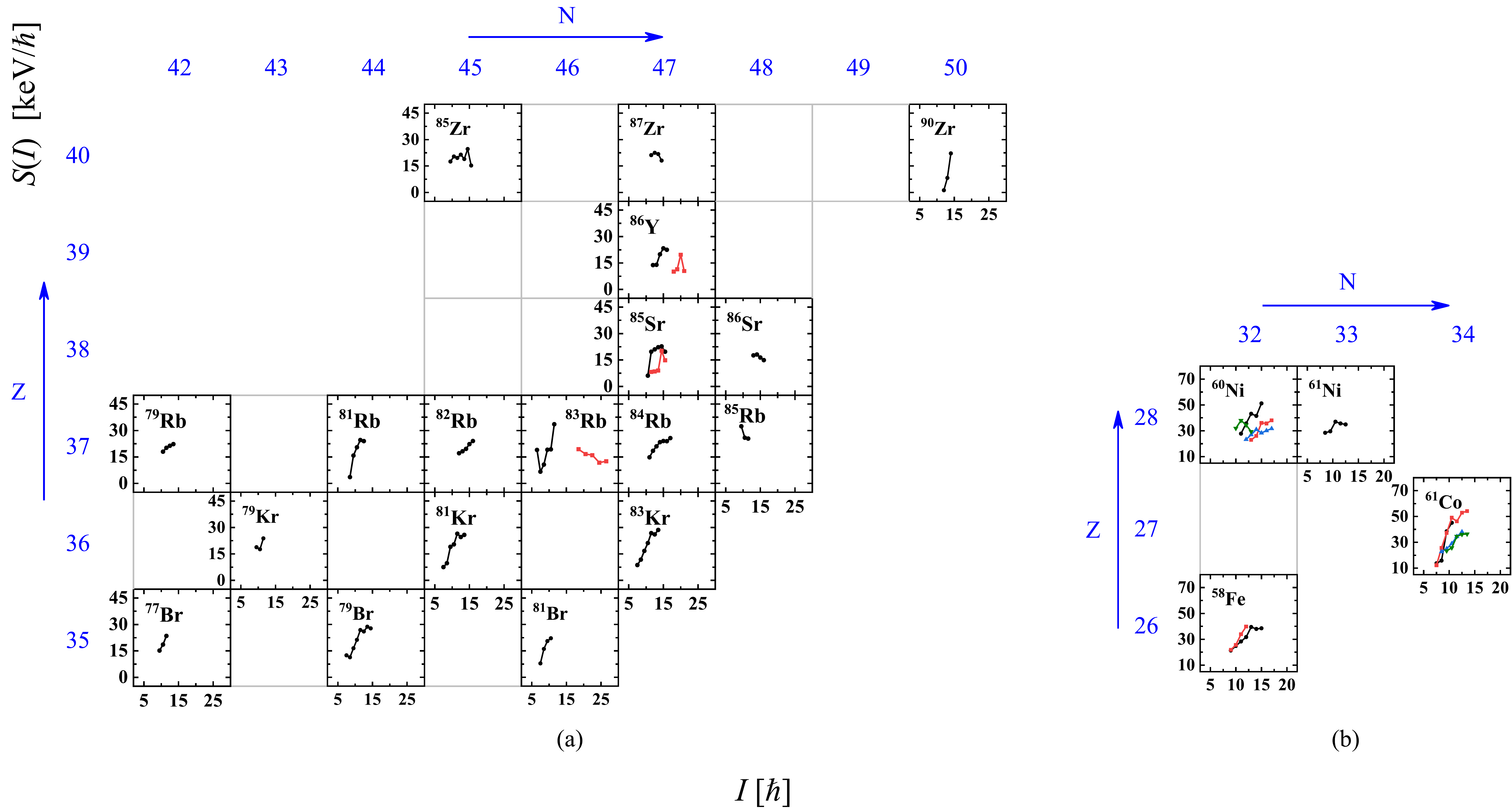}
		\caption{(Color online) Energy staggering parameter as functions of spin for magnetic rotational bands in $A$ $\sim$ (a) 80 and (b) 60 mass regions.}
		\label{13}
	\end{figure}

\begin{figure}[H]
		\centering
		\includegraphics[width=15cm]{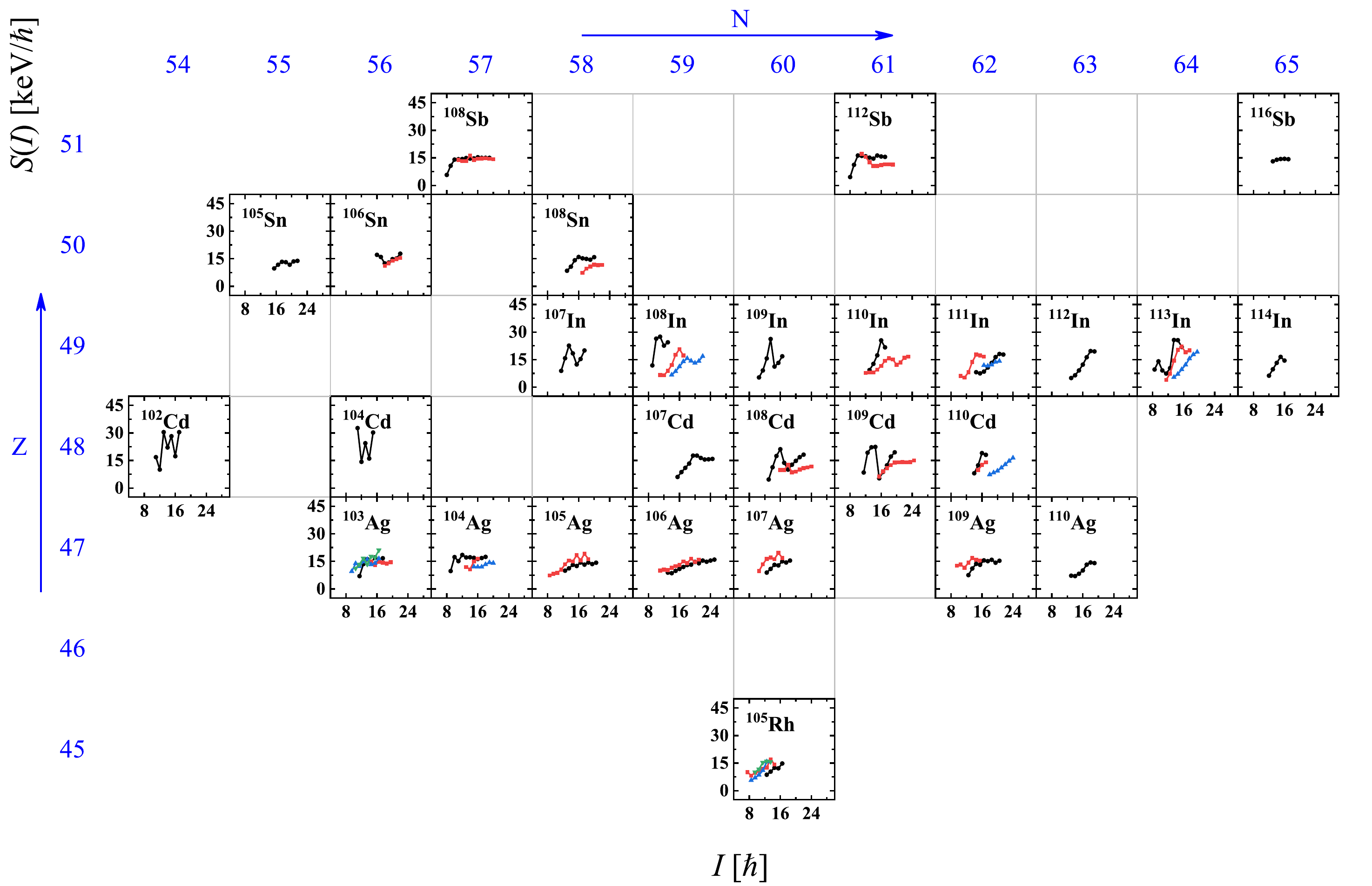}
		\caption{(Color online) Energy staggering parameter as functions of spin for magnetic rotational bands in $A$ $\sim$ 110 mass region.}
		\label{14}
	\end{figure}

\begin{figure}[H]
		\centering
		\includegraphics[width=18cm]{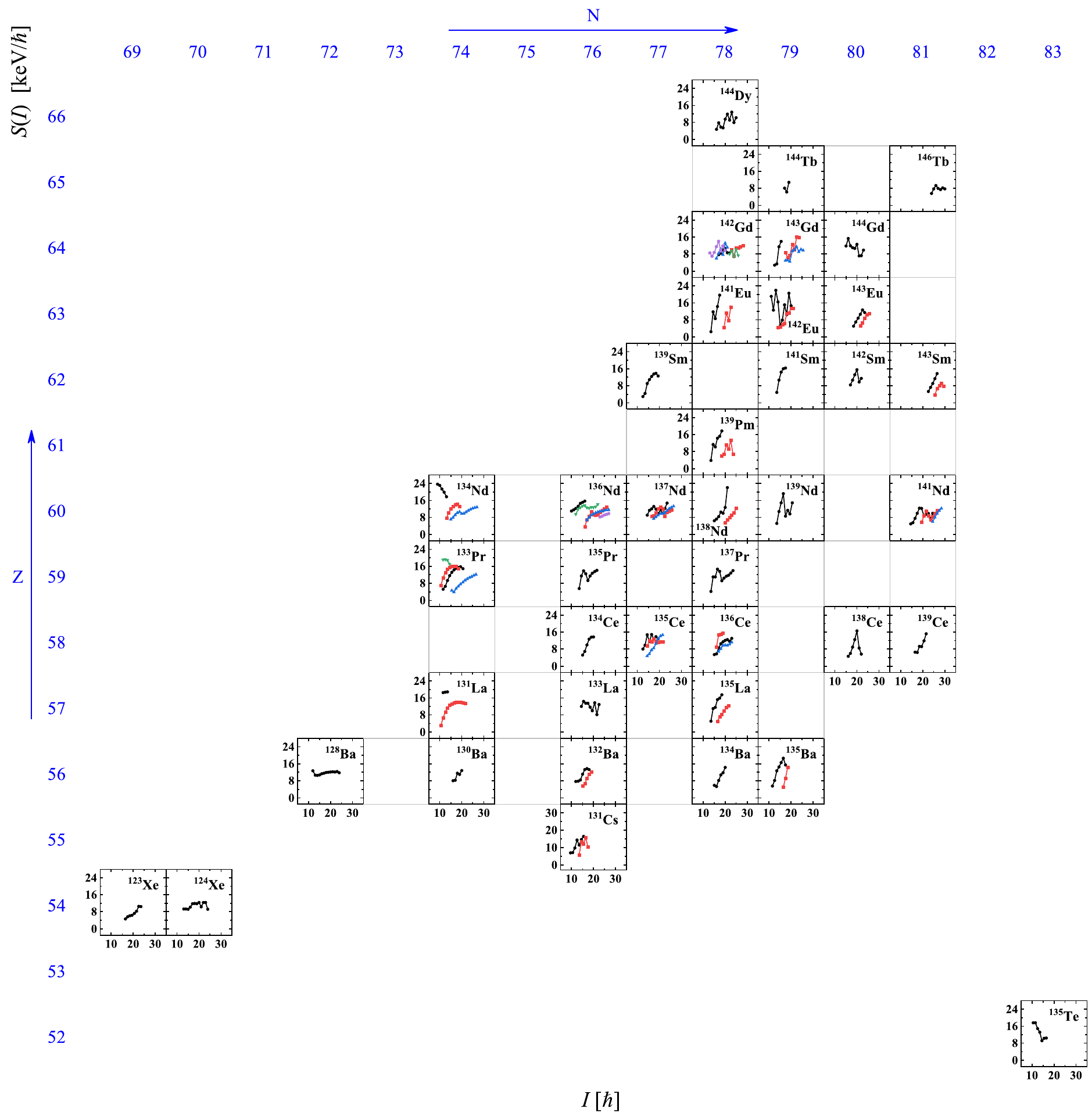}
		\caption{(Color online) Energy staggering parameter as functions of spin for magnetic rotational bands in $A$ $\sim$ 140 mass region.}
		\label{15}
	\end{figure}

\begin{figure}[H]
		\centering
		\includegraphics[width=16.5cm]{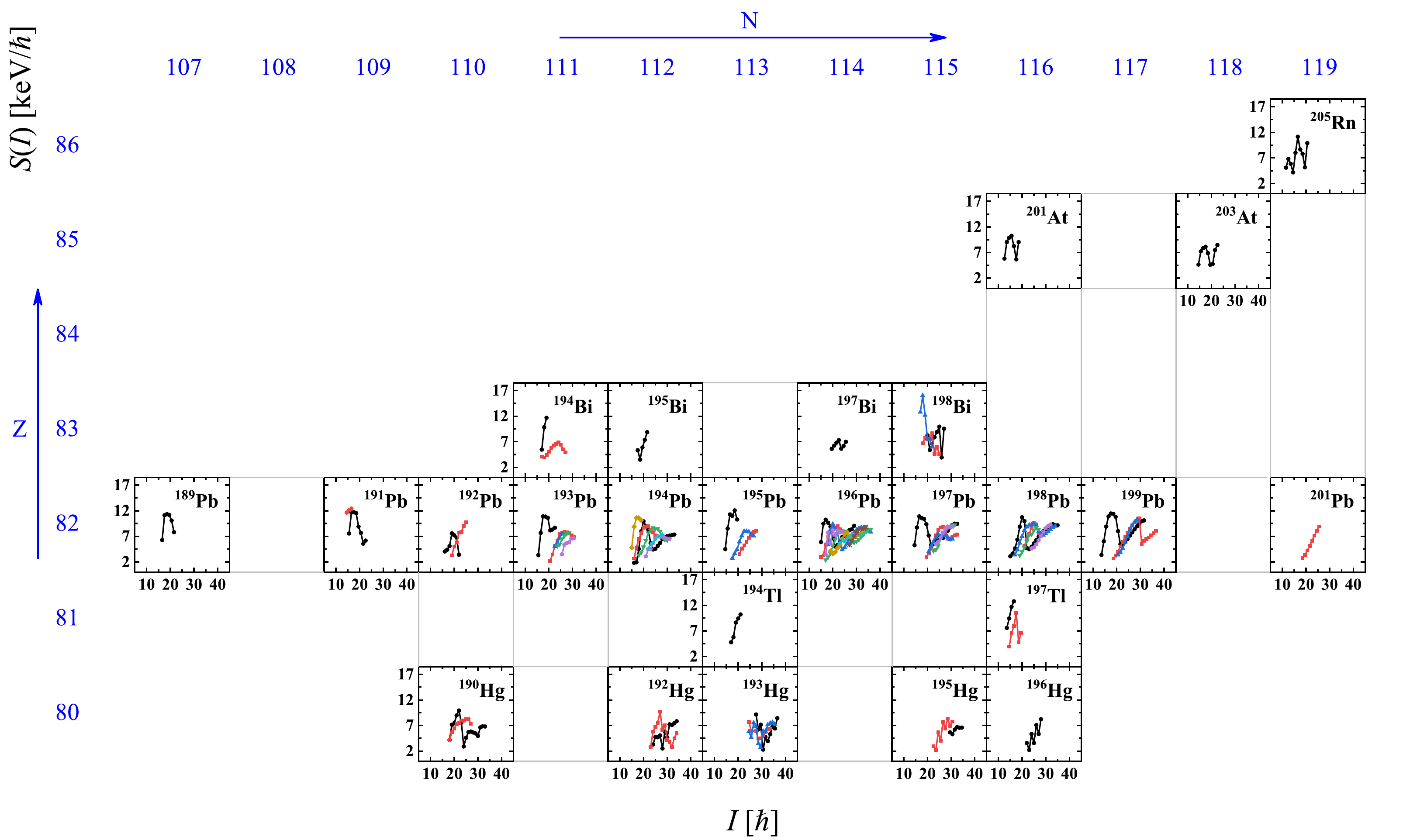}
		\caption{(Color online) Energy staggering parameter as functions of spin for magnetic rotational bands in $A$ $\sim$ 190 mass region.}
		\label{16}
	\end{figure}

	\cref{13,14,15,16} show the plots of $S(I)$ vs. $I$ for all magnetic rotational bands in $A$ $\sim$ 80, 110, 140, and 190 mass regions, respectively. No signature splitting should be observed in ideal MR bands due to the pure individual motion of nucleons \cite{RN425,RN429,RNS99}. However, the phenomena of signature splitting are shown in many bands, such as in $^{105,~109}$Ag, $^{102,~104}$Cd, $^{135}$Ce, $^{142}$Eu, $^{144}$Dy, $^{193,~195,~196}$Hg, and $^{198}$Bi (No judgement on the presence or absence of splitting could however be made in many cases where only 4$-$5 levels have been observed.)
	Moreover, it can be seen that $S(I)$ in most MR bands increase as the spin increasing except for the band-crossing regions. But $S(I)$ shows decreasing trend as a function of spin in $^{83,~85}$Rb, $^{86}$Sr, $^{135}$Te, $^{133}$Pr, $^{134}$Nd, $^{144}$Gd, $^{191}$Pb and $^{198}$Bi. In addition, we can infer from $\mathcal{J}$$^{(1)}$$(=1/[2S(I)])$ that both the upper and lower bounds of the $S(I)$ values decrease with the change of the mass region from $A$ $\sim$ 60 to $A$ $\sim$ 190.

	\subsection{The magnetic dipole reduced transition probability vs. spin}
\begin{figure}[H]
		\centering
		\includegraphics[width=8.6cm]{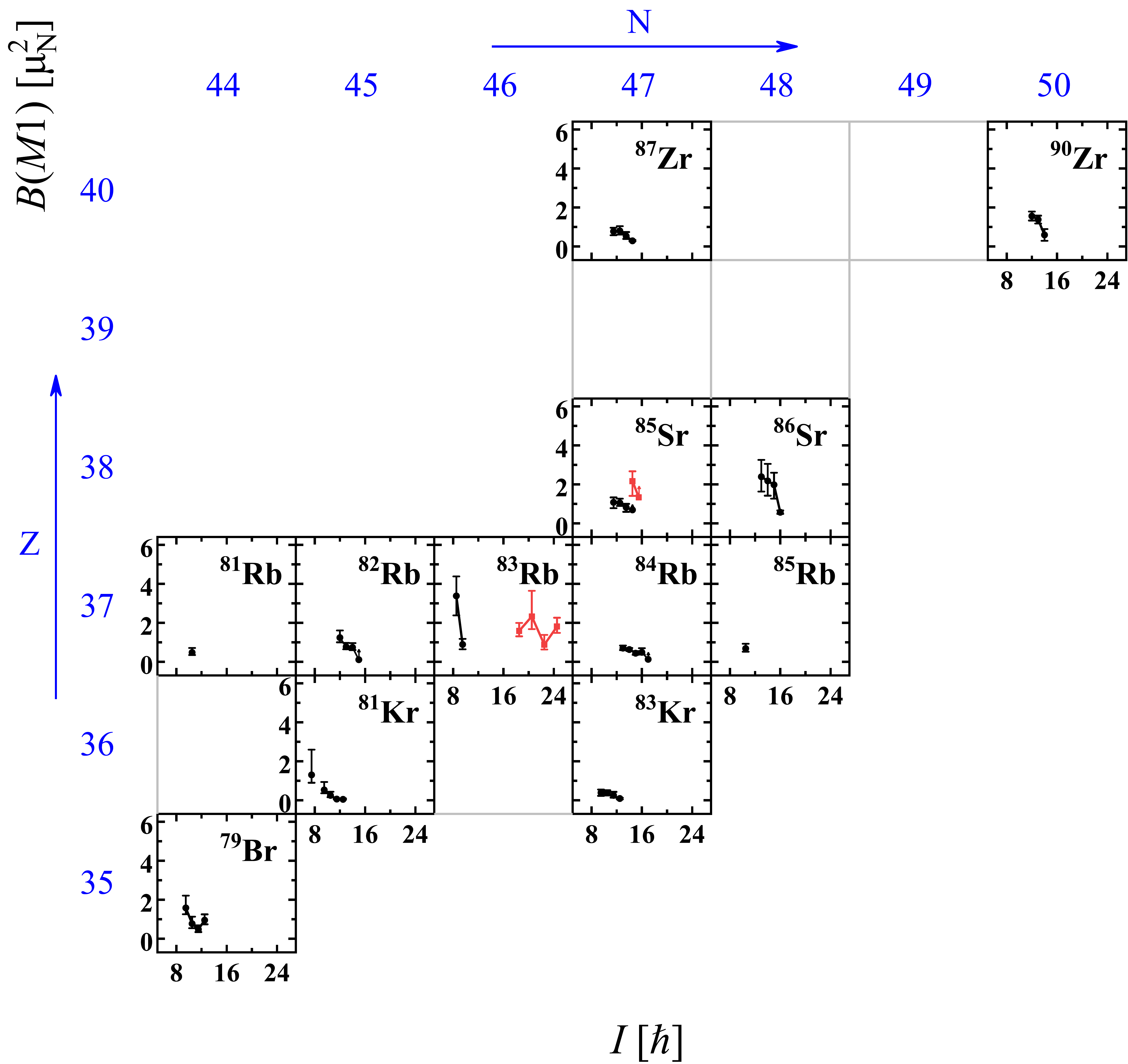}
		\caption{(Color online) Magnetic dipole reduced transition probability versus spin for magnetic rotational bands in $A$ $\sim$ 80 mass region.}
		\label{17}
	\end{figure}

\begin{figure}[H]
		\centering
		\includegraphics[width=12.5cm]{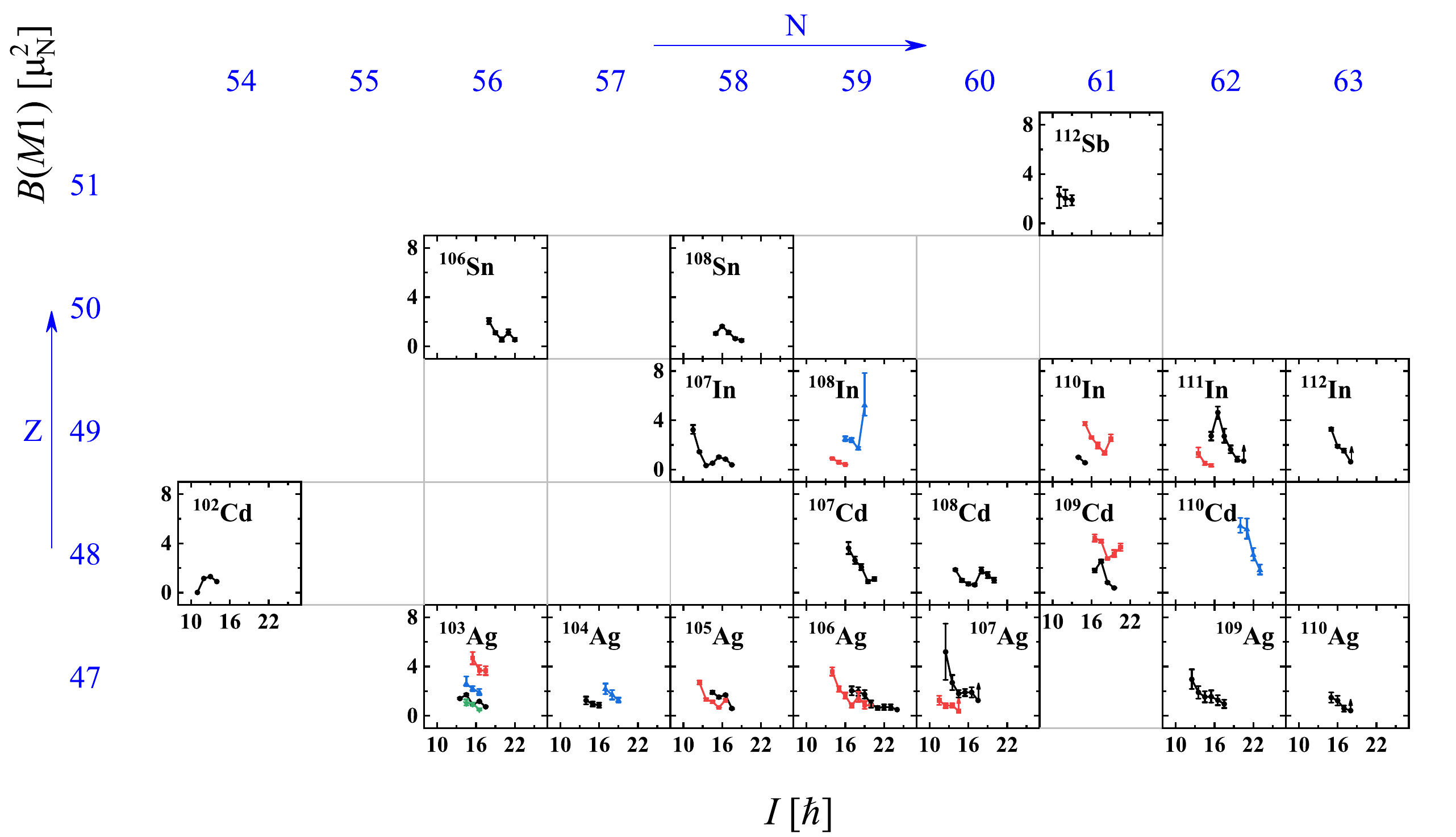}
		\caption{(Color online) Magnetic dipole reduced transition probability versus spin for magnetic rotational bands in $A$ $\sim$ 110 mass region.}
		\label{18}
	\end{figure}

\begin{figure}[H]
		\centering
		\includegraphics[width=15.5cm]{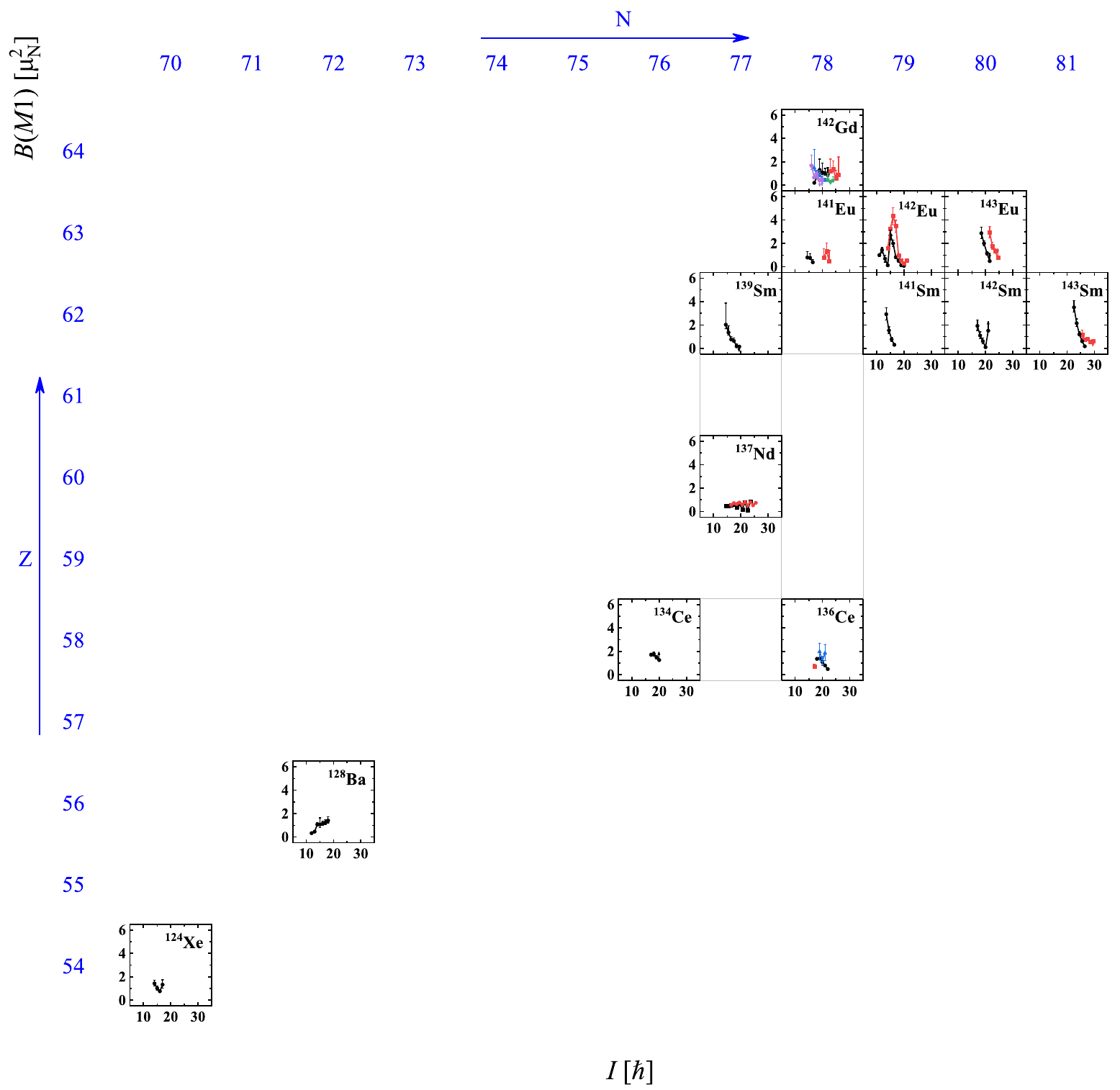}
		\caption{(Color online) Magnetic dipole reduced transition probability versus spin for magnetic rotational bands in $A$ $\sim$ 140 mass region.}
		\label{19}
	\end{figure}

\begin{figure}[H]
		\centering
		\includegraphics[width=9.2cm]{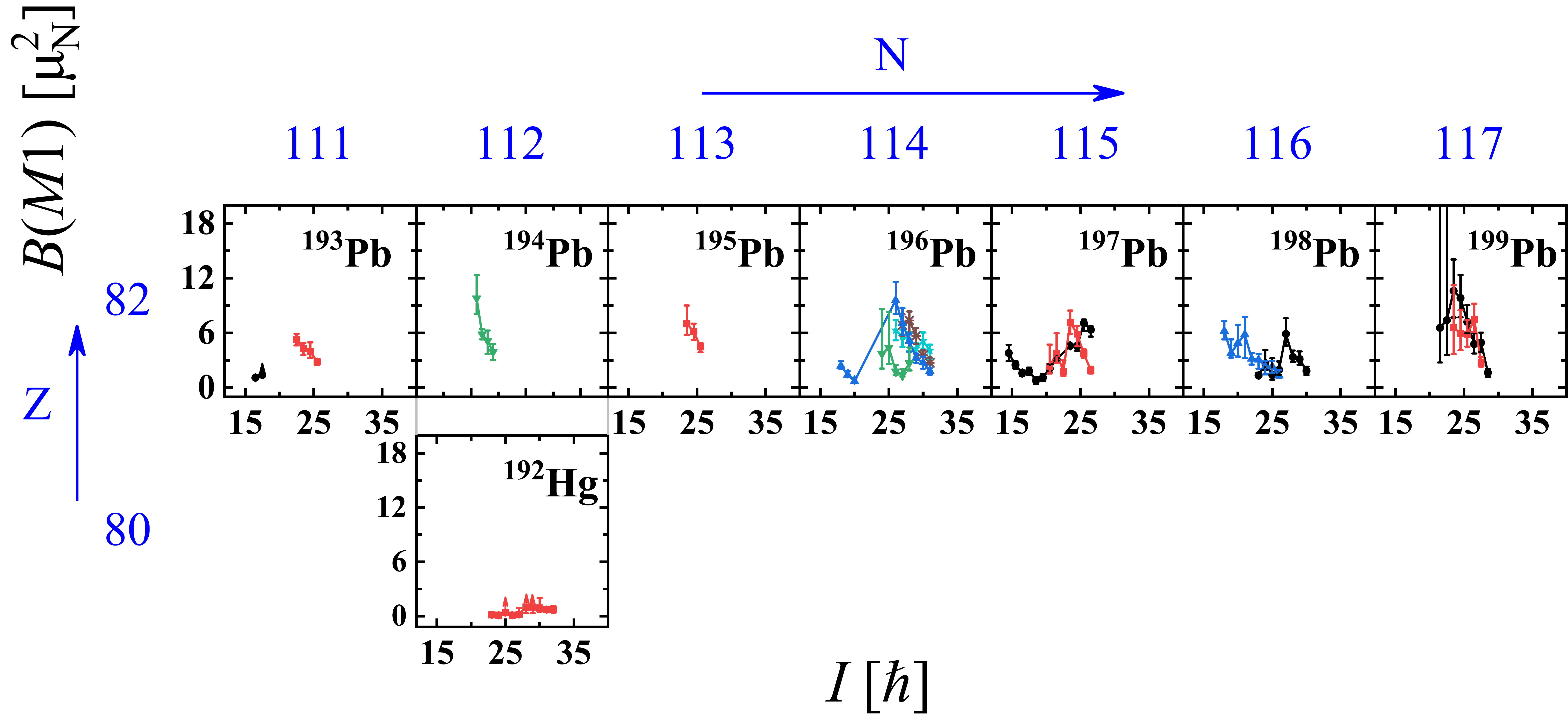}
		\caption{(Color online) Magnetic dipole reduced transition probability versus spin for magnetic rotational bands in $A$ $\sim$ 190 mass region.}
		\label{20}
	\end{figure}

	The plots of $B(M1)$ vs. $I$ for all magnetic rotational bands in $A$ $\sim$ 80, 110, 140, and 190 mass regions are given in \cref{17,18,19,20}, respectively. The $B(M1)$ values of the magnetic rotation are large (of the order of several $\mu_{N}^{2}$ units) and the $B(M1)$ values decrease with the increasing spin. However, it should be noted that $B(M1)$ values show a large staggering in the MR bands of $^{83}$Rb, $^{103}$Ag, $^{137}$Nd, and $^{197}$Pb. 
	
	\subsection{The electric quadrupole reduced transition probability vs. spin}
\begin{figure}[H]
		\centering
		\includegraphics[width=5.8cm]{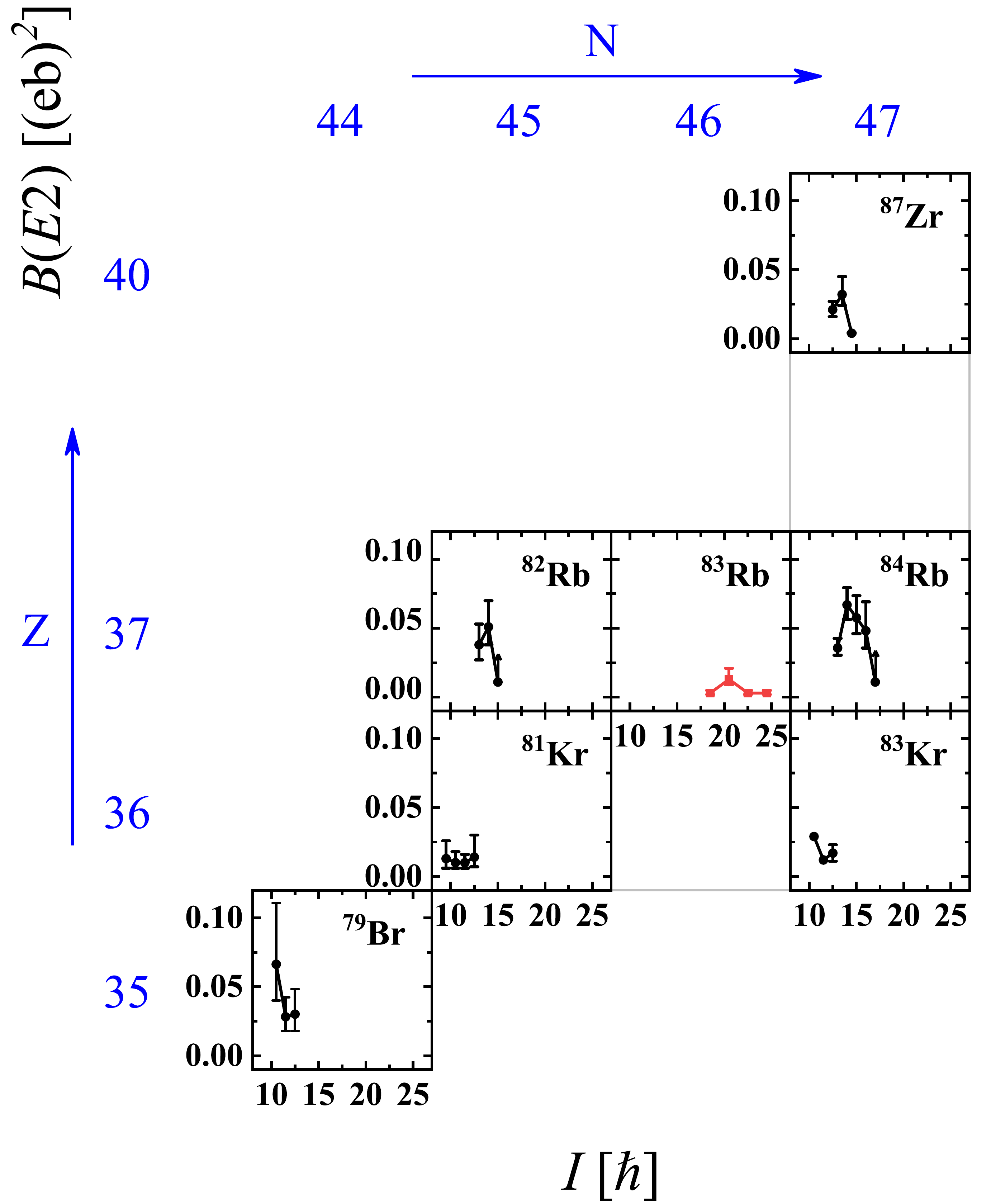}
		\caption{(Color online) Electric quadrupole reduced transition probability versus spin for magnetic rotational bands in $A$ $\sim$ 80 mass region.}
		\label{21}
	\end{figure}

\begin{figure}[H]
		\centering
		\includegraphics[width=9.3cm]{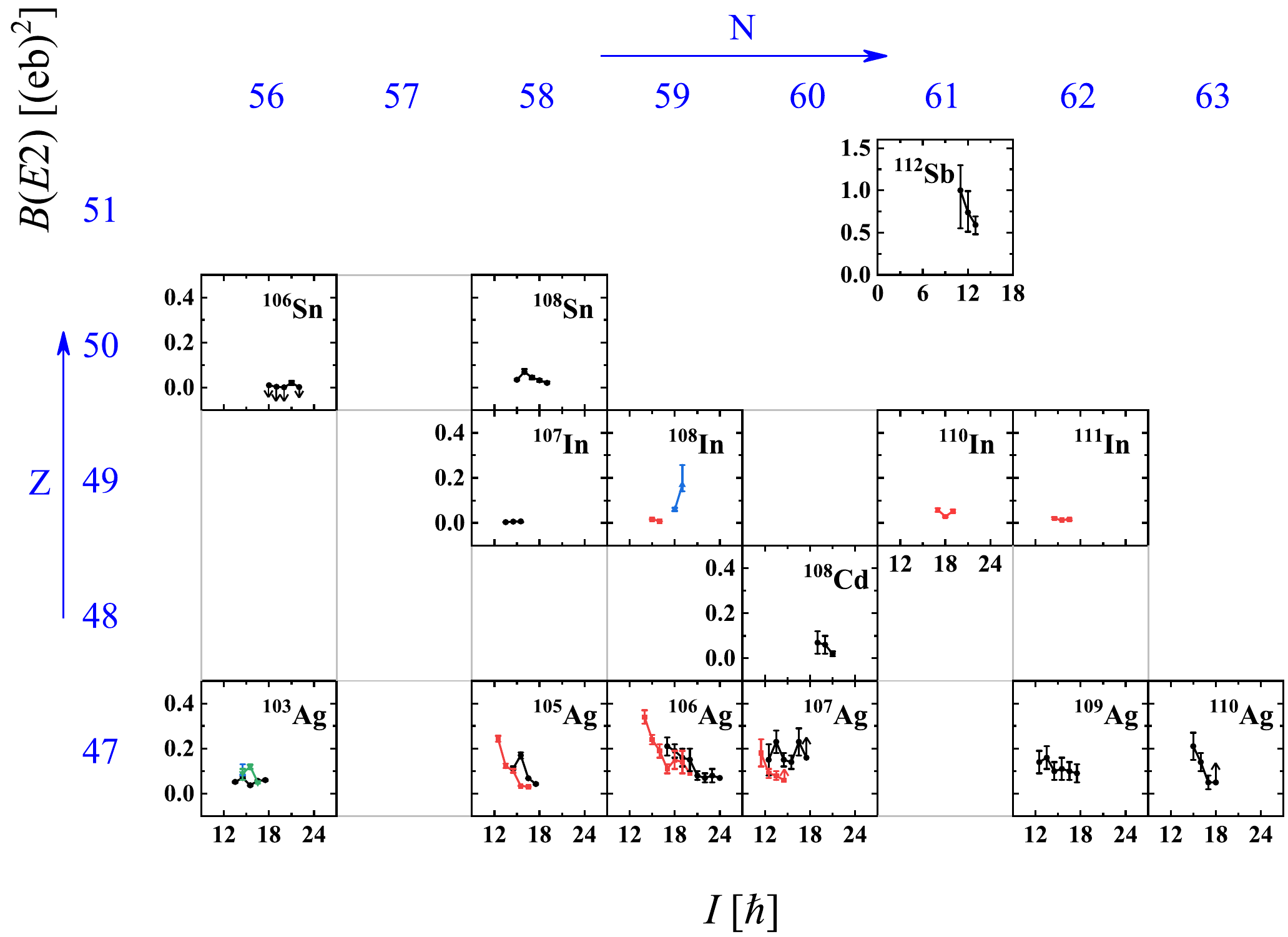}
		\caption{(Color online) Electric quadrupole reduced transition probability versus spin for magnetic rotational bands in $A$ $\sim$ 110 mass region.}
		\label{22}
	\end{figure}

\begin{figure}[H]
		\centering
		\includegraphics[width=15.2cm]{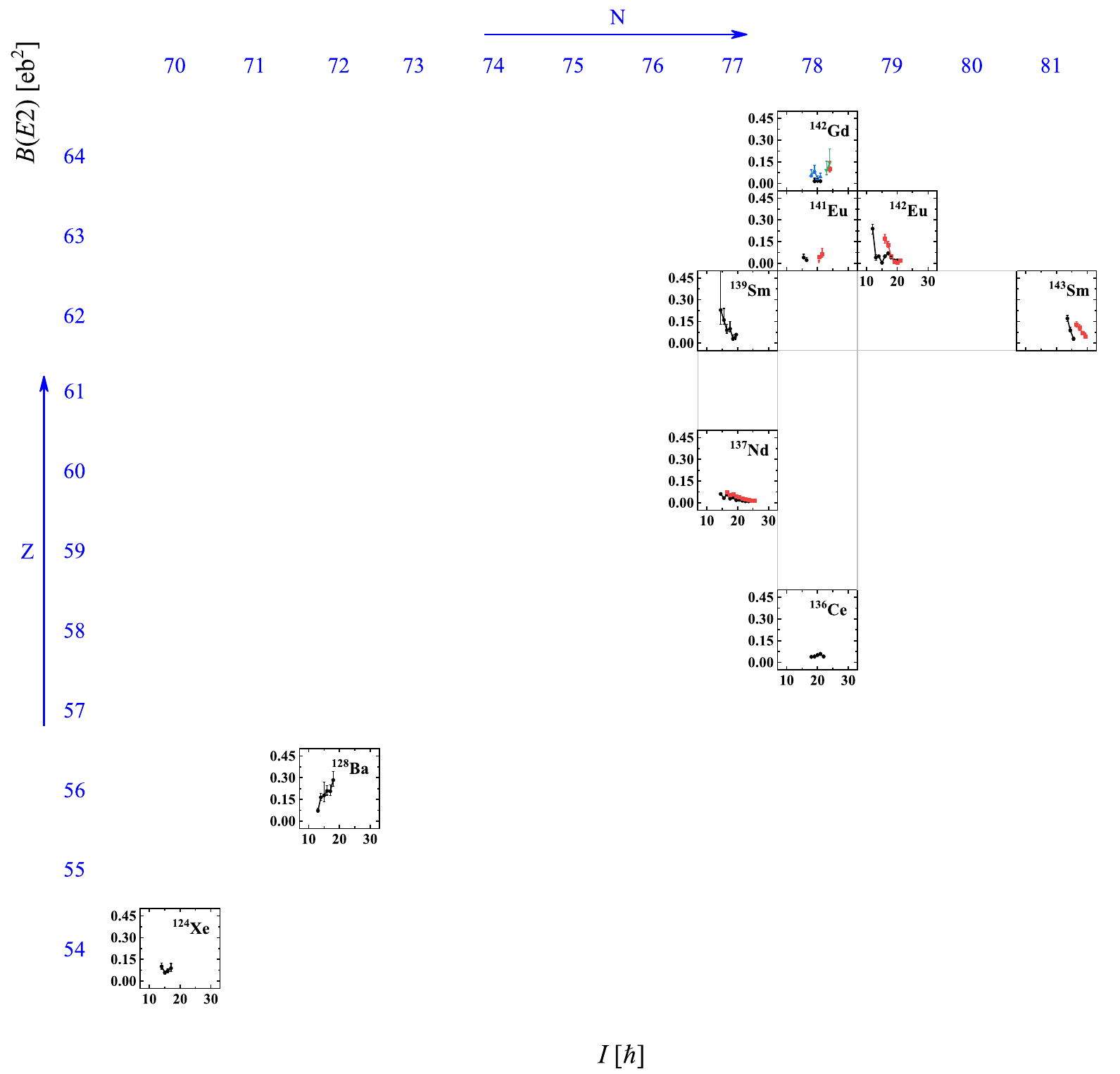}
		\caption{(Color online) Electric quadrupole reduced transition probability versus spin for magnetic rotational bands in $A$ $\sim$ 140 mass region.}
		\label{23}
	\end{figure}

\begin{figure}[H]
		\centering
		\includegraphics[width=9cm]{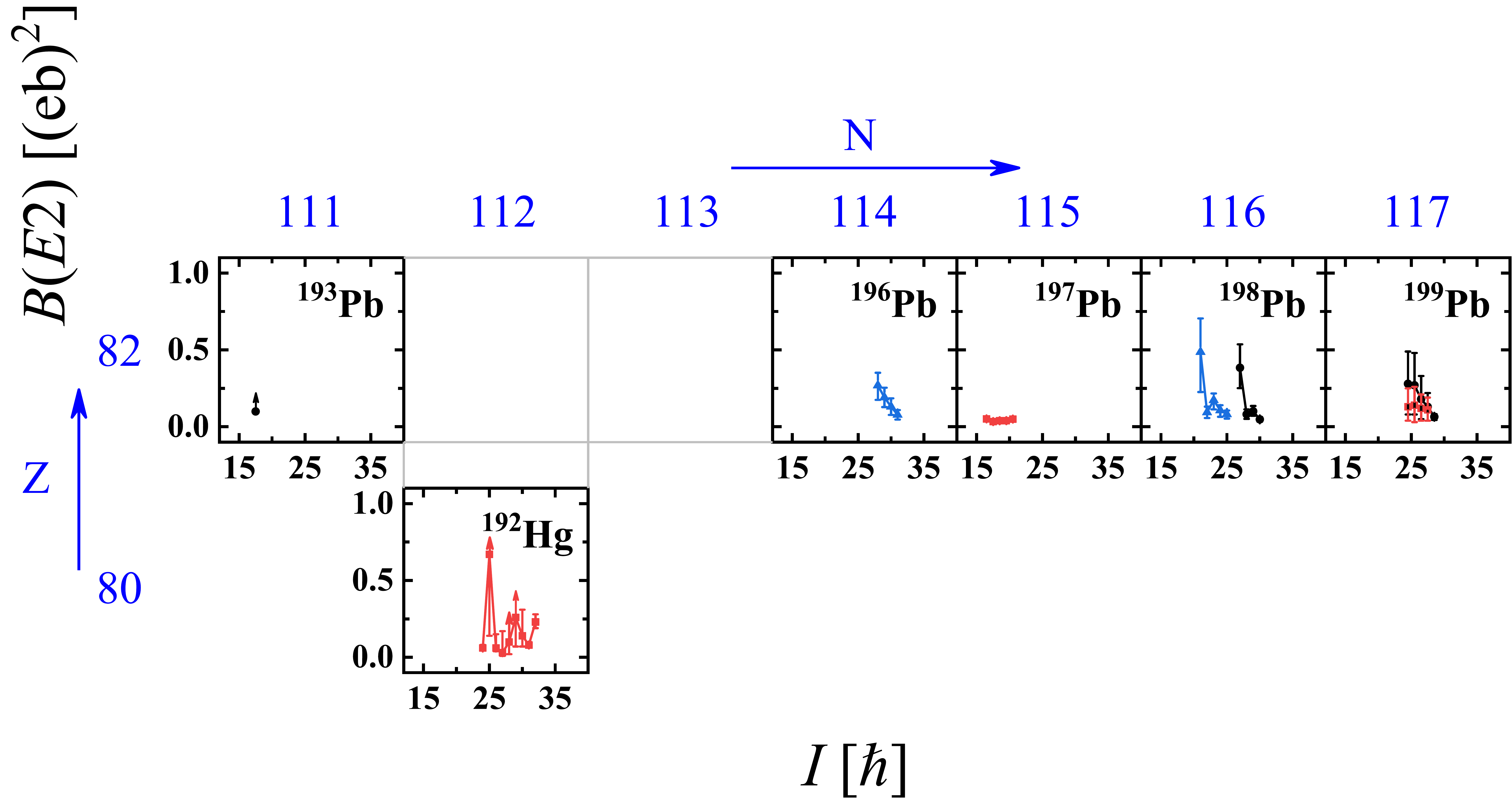}
		\caption{(Color online) Electric quadrupole reduced transition probability versus spin for magnetic rotational bands in $A$ $\sim$ 190 mass region.}
		\label{24}
	\end{figure}

	\cref{21,22,23,24} show the electric quadrupole reduced transition probabilities $B(E2)$ for magnetic rotational bands in $A$ $\sim$ 80, 110, 140, and 190 mass regions, respectively. It is well known that the values of $B(E2)$ are small, $B(E2)$ $\sim$ 0.1 (eb)$^{2}$ \cite{RN93}. Indeed, most nuclei clearly satisfy the criterion. However, the values of $B(E2)$ are less than 0.1 (eb)$^{2}$ in $A$ $\sim$ 80 mass region, while the values of $B(E2)$ vary between 0.5$-$1.0 (eb)$^{2}$ in $^{112}$Sb. Furthermore, $B(E2)$ decreases as the spin increases except for $^{108}$In, $^{128}$Ba, $^{141}$Eu, and $^{142}$Gd. $B(E2)$ values also show a large staggering in the MR bands of $^{107}$Ag, $^{137}$Nd, and $^{192}$Hg.

	\subsection{The $B(M1)/B(E2)$ ratio vs. spin }
\begin{figure}[H]
		\centering
		\includegraphics[width=7.2cm]{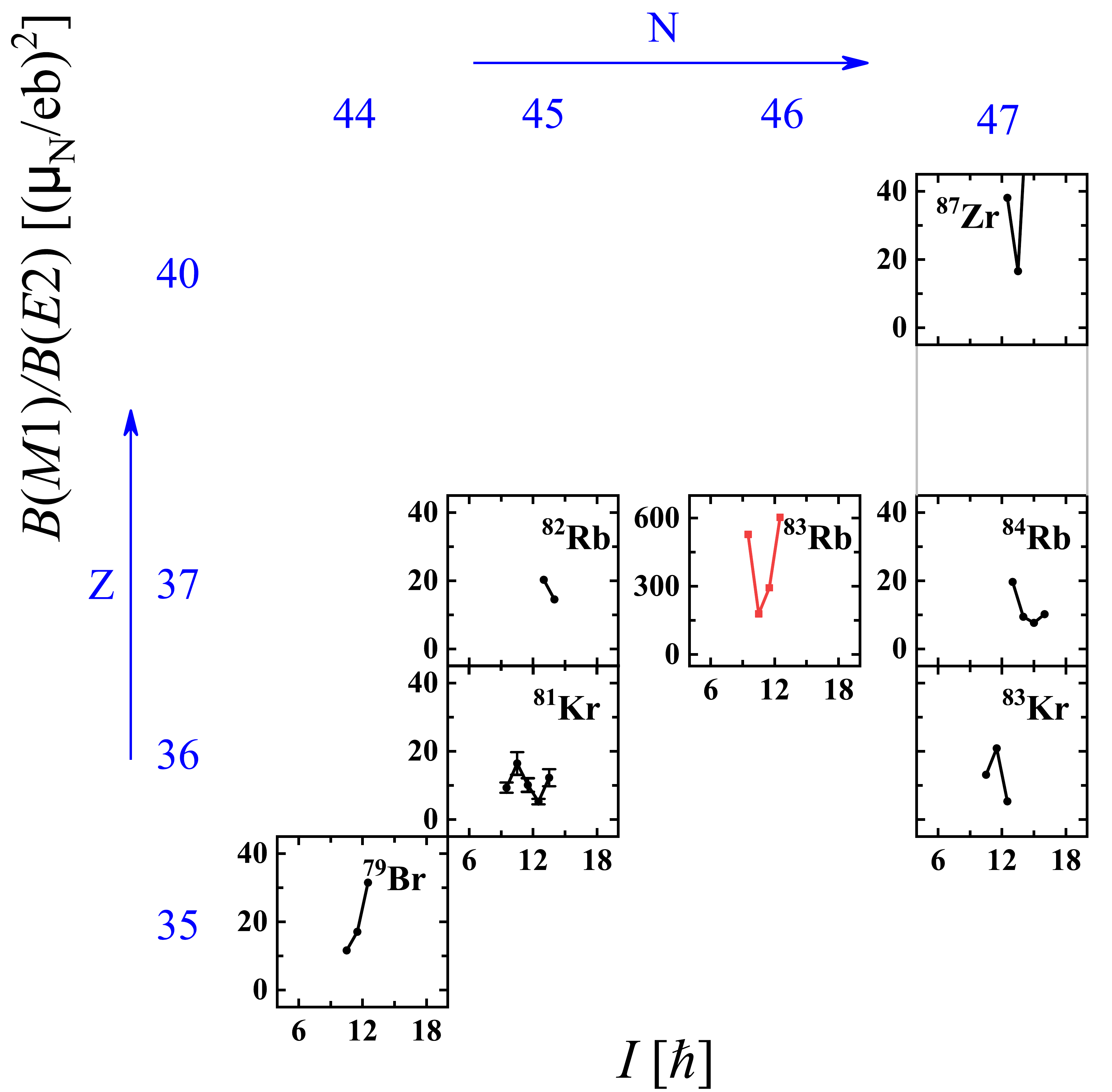}
		\caption{(Color online) $B(M1)/B(E2)$ ratio versus spin for magnetic rotational bands in $A$ $\sim$ 80 mass region.}
		\label{25}
	\end{figure}

\begin{figure}[H]
		\centering
		\includegraphics[width=11.9cm]{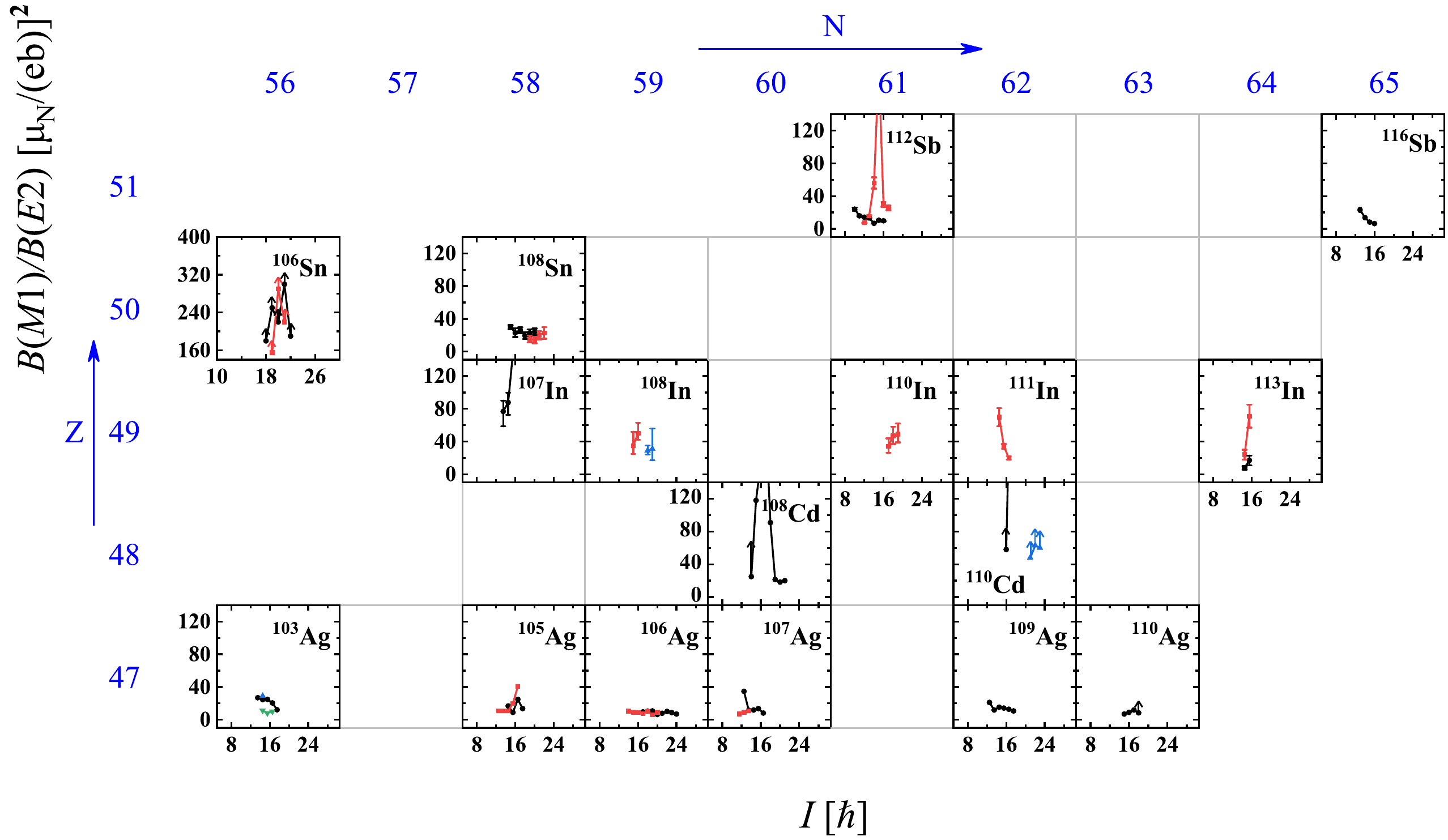}
		\caption{(Color online) $B(M1)/B(E2)$ ratio versus spin for magnetic rotational bands in $A$ $\sim$ 110 mass region.}
		\label{26}
	\end{figure}

\begin{figure}[H]
		\centering
		\includegraphics[width=15.1cm]{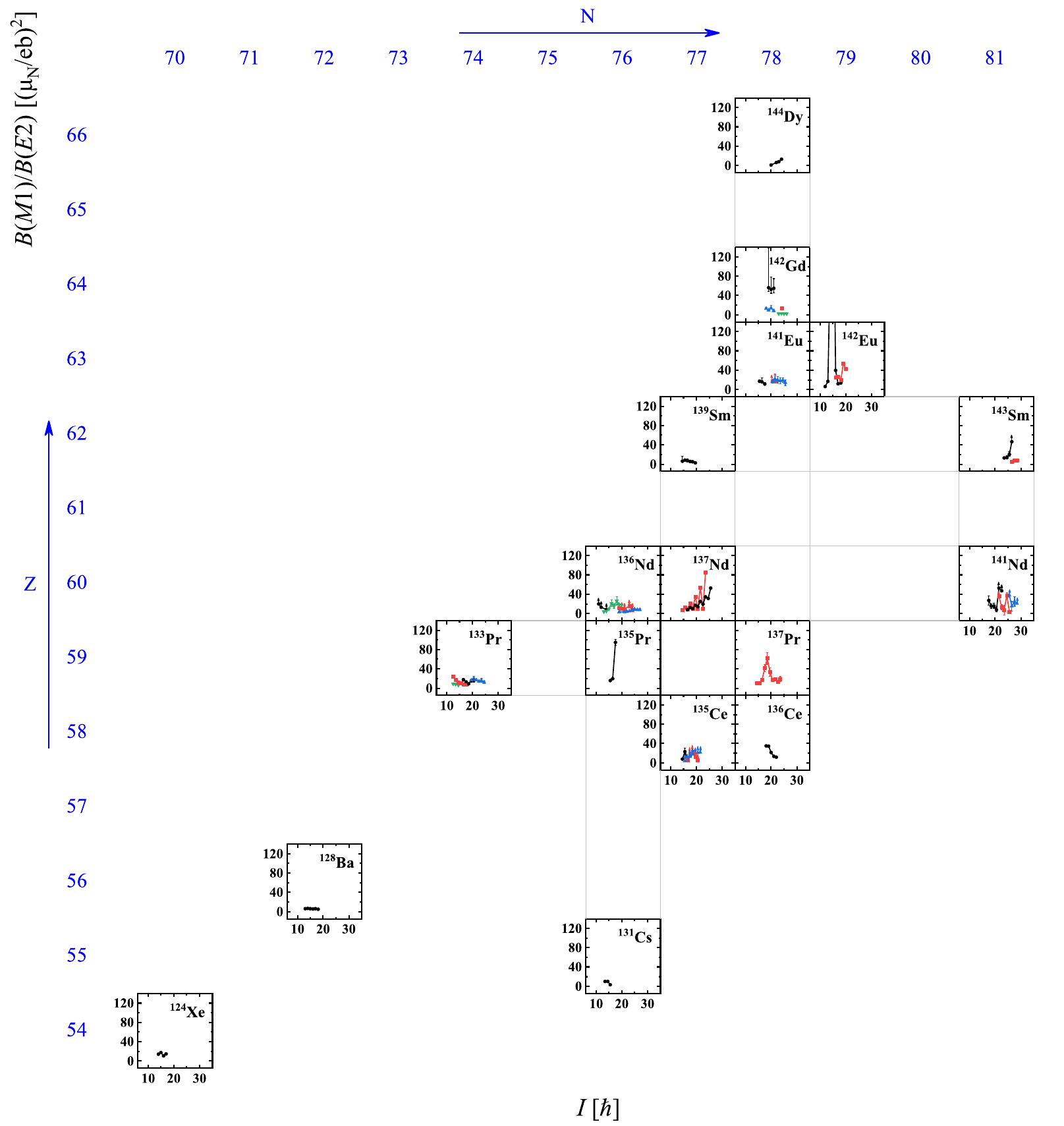}
		\caption{(Color online) $B(M1)/B(E2)$ ratio versus spin for magnetic rotational bands in $A$ $\sim$ 140 mass region.}
		\label{27}
	\end{figure}

\begin{figure}[H]
		\centering
		\includegraphics[width=14cm]{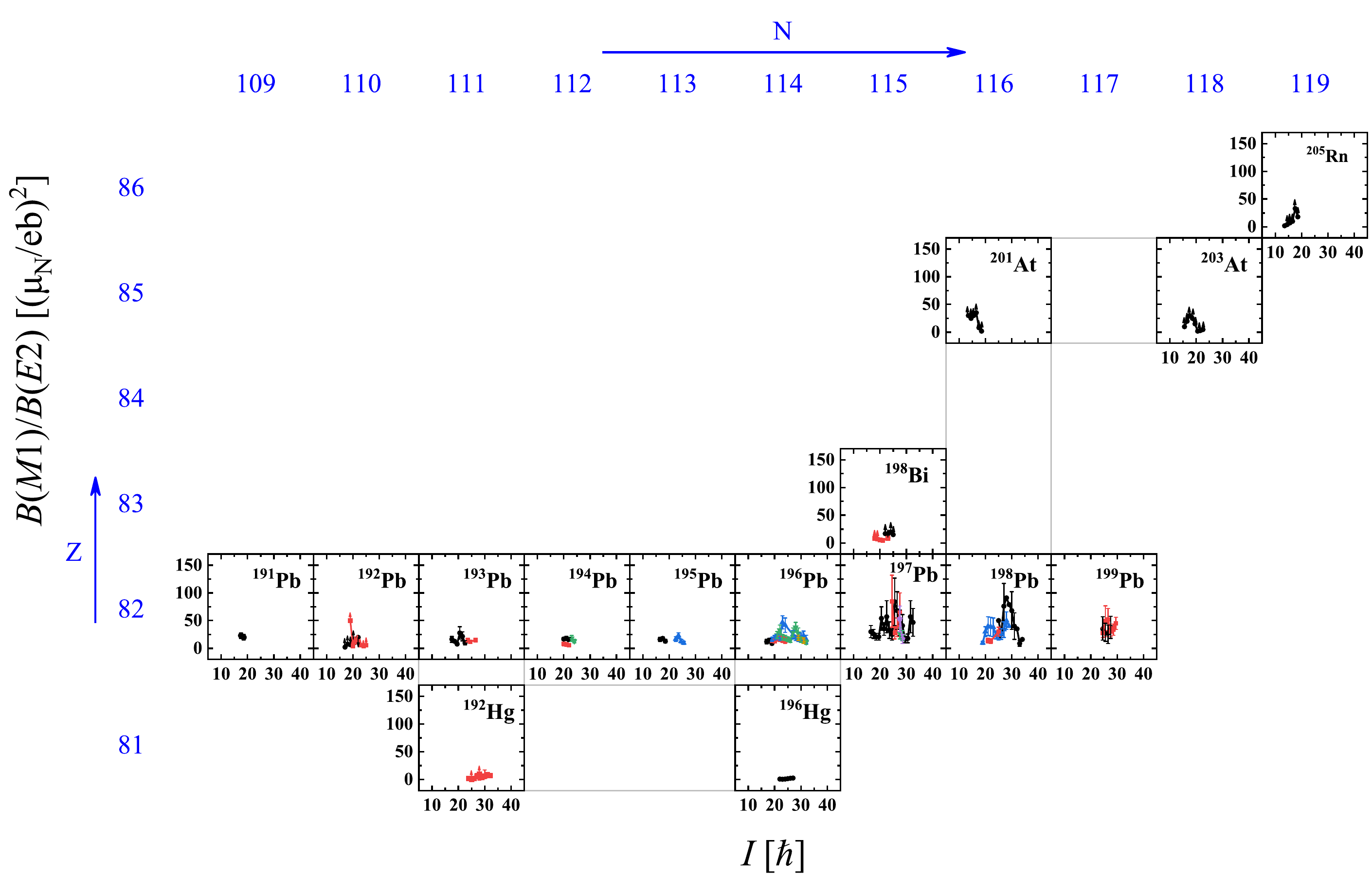}
		\caption{(Color online) $B(M1)/B(E2)$ ratio versus spin for magnetic rotational bands in $A$ $\sim$ 190 mass region.}
		\label{28}
	\end{figure}

	The $B(M1)/B(E2)$ ratios for all magnetic rotational bands in $A$ $\sim$ 80, 110, 140, and 190 mass regions are shown in \cref{25,26,27,28}, respectively. For magnetic rotational bands, the $B(M1)/B(E2)$ values are generally more than 20 [$\mu_{N}$/(eb)]$^{2}$ \cite{RN93}. However, some values of MR bands are around 10 [$\mu_{N}$/(eb)]$^{2}$ in $^{106,~107,~109}$Ag, $^{112}$Sb, $^{128}$Ba, $^{139}$Sm, $^{142}$Gd, $^{144}$Dy, $^{192,~196}$Hg, and $^{192,~193,~194}$Pb, smaller than the criterion for magnetic rotation mentioned above. In contrast, the $B(M1)/B(E2)$ values are large in some MR bands, for example in $^{83}$Rb whose values of $B(M1)/B(E2)$ are around 88$-$603 [$\mu_{N}$/(eb)]$^{2}$, and more than 150 [$\mu_{N}$/(eb)]$^{2}$ in $^{106}$Sn. It is worthwhile to mention here that there is a significant staggering of $B(M1)/B(E2)$ values observed in $^{106}$Sn, $^{137}$Nd, and $^{197}$Pb.
	
	Up to now, the MR bands have been shown in 64 odd-$A$ nuclei, 25 odd-odd nuclei, and 34 even-even nuclei. It is obvious that the MR bands are reported in the odd-odd isotopes of Sb and Tb as well as the odd-$A$ isotopes of Br, Kr, La, and Pr. According to the systematics of magnetic rotation, we may infer that $^{80}$Rb, $^{108}$Ag, $^{137}$Ce, $^{135,~140}$Nd, $^{140}$Sm, $^{190}$Pb, and $^{196,~201}$Bi could be the candidate nuclei with MR bands. 
	
	\section{Systematics of antimagnetic rotational bands}
	
	\subsection{The spin vs. rotational frequency}
\begin{figure}[H]
		\centering
		\includegraphics[width=17.3cm]{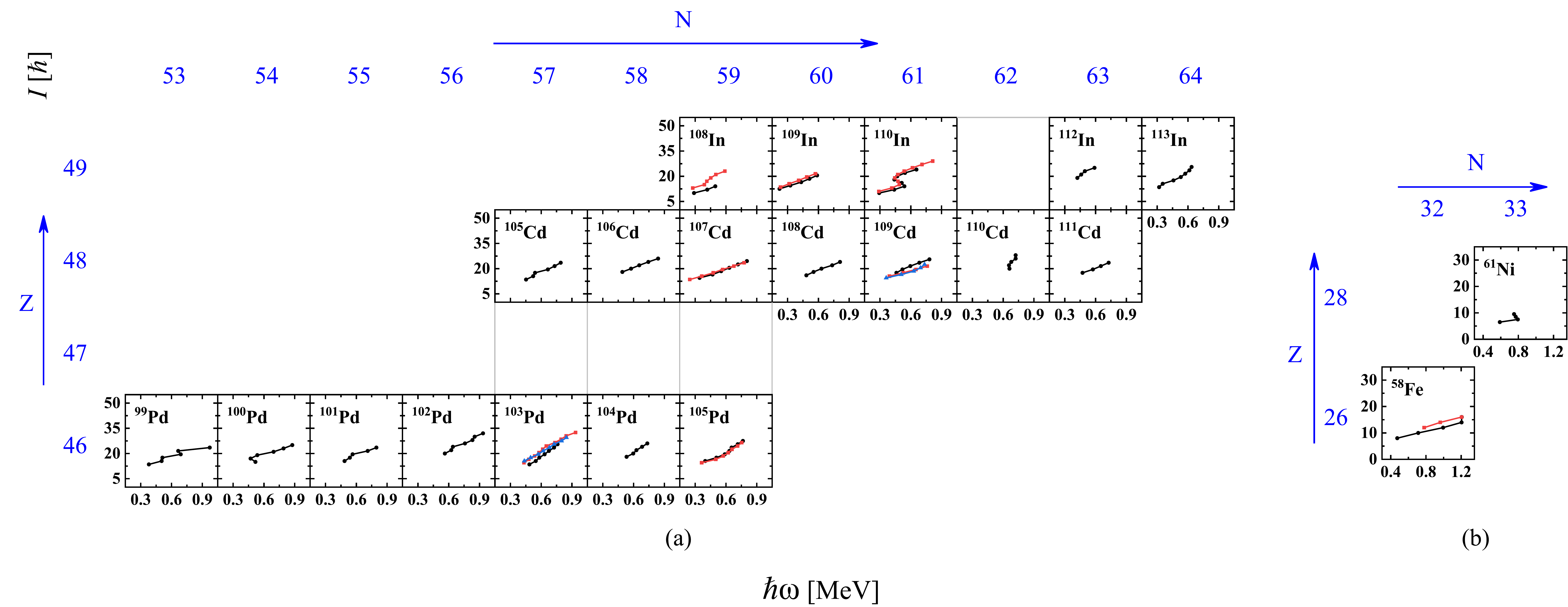}
		\caption{(Color online) Spin versus rotational frequency for antimagnetic rotational bands in $A$ $\sim$ (a) 110 and (b) 60 mass regions.}
		\label{33}
	\end{figure}

\begin{figure}[H]
		\centering
		\includegraphics[width=8cm]{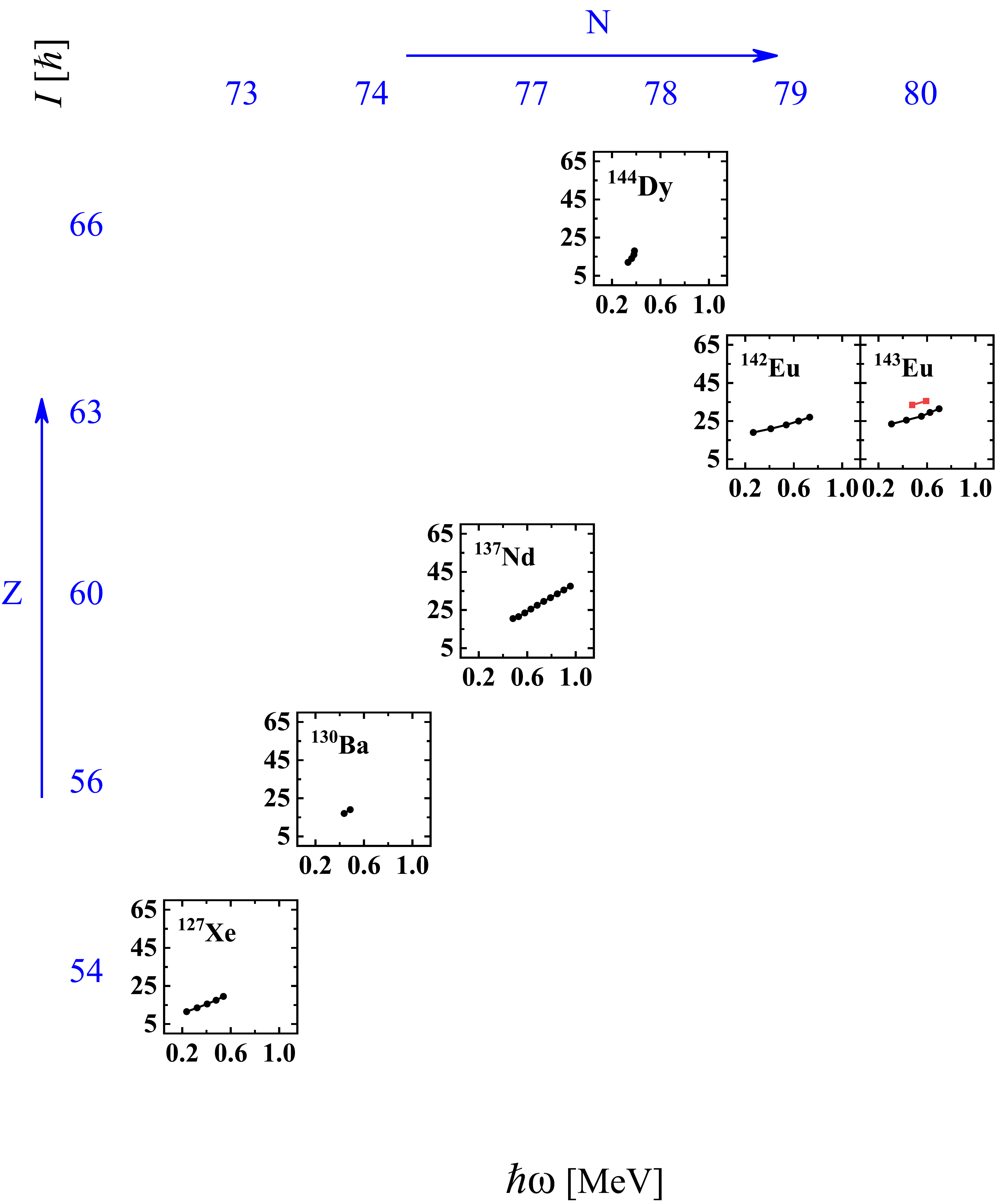}
		\caption{(Color online) Spin versus rotational frequency for antimagnetic rotational bands in $A$ $\sim$ 140 mass region.}
		\label{34}
	\end{figure}

	\cref{33,34} show the relations between the spins and rotational frequencies for antimagnetic rotational bands in $A$ $\sim$ 60, 110, and 140 mass regions, respectively. The spin generally grows steadily as the rotational frequency increases. It is obvious that the band in $^{99}$Pd shows a significant splitting phenomenon.   

	\subsection{The kinematic moment of inertia vs. rotational frequency}
\begin{figure}[H]
		\centering
		\includegraphics[width=17.3cm]{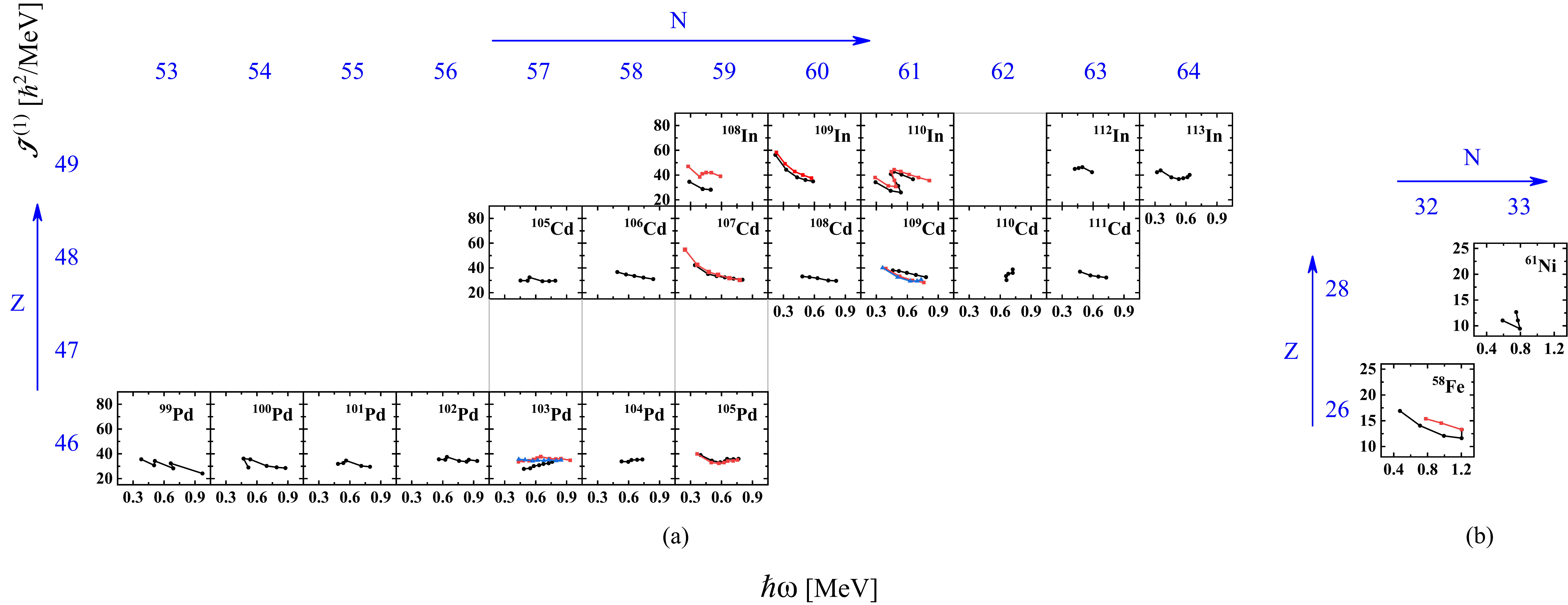}
		\caption{(Color online) Kinematic moment of inertia versus rotational frequency for antimagnetic rotational bands in $A$ $\sim$ (a) 110 and (b) 60 mass regions.}
		\label{29}
	\end{figure}

\begin{figure}[H]
		\centering
		\includegraphics[width=8cm]{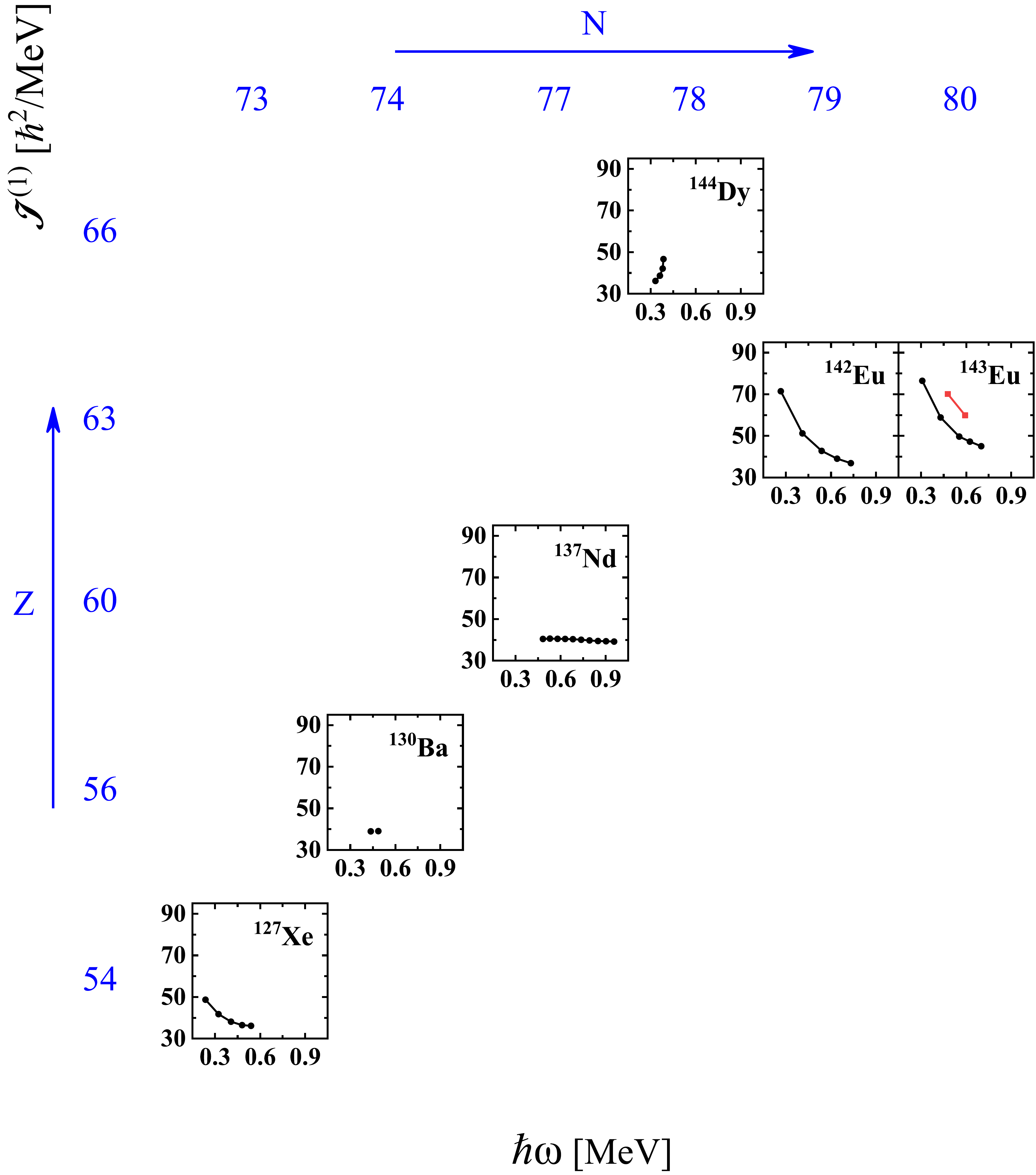}
		\caption{(Color online) Kinematic moment of inertia versus rotational frequency for antimagnetic rotational bands in $A$ $\sim$ 140 mass region.}
		\label{30}
	\end{figure}

	The kinematic moments of inertia $\mathcal{J}$$^{(1)}$ for all antimagnetic rotational bands in $A$ $\sim$ 60, 110 and 140 mass regions are given in \cref{29,30}, respectively. Generally, the kinematic moment of inertia decreases as the rotational frequency increases. However, there is an abnormal phenomenon in $^{144}$Dy where $\mathcal{J}$$^{(1)}$ significantly increases with the rotational frequency. The band in $^{99}$Pd also shows an obvious splitting phenomenon.
	\subsection{The dynamic moments of inertia vs. rotational frequency}
\begin{figure}[H]
		\centering
		\includegraphics[width=17.3cm]{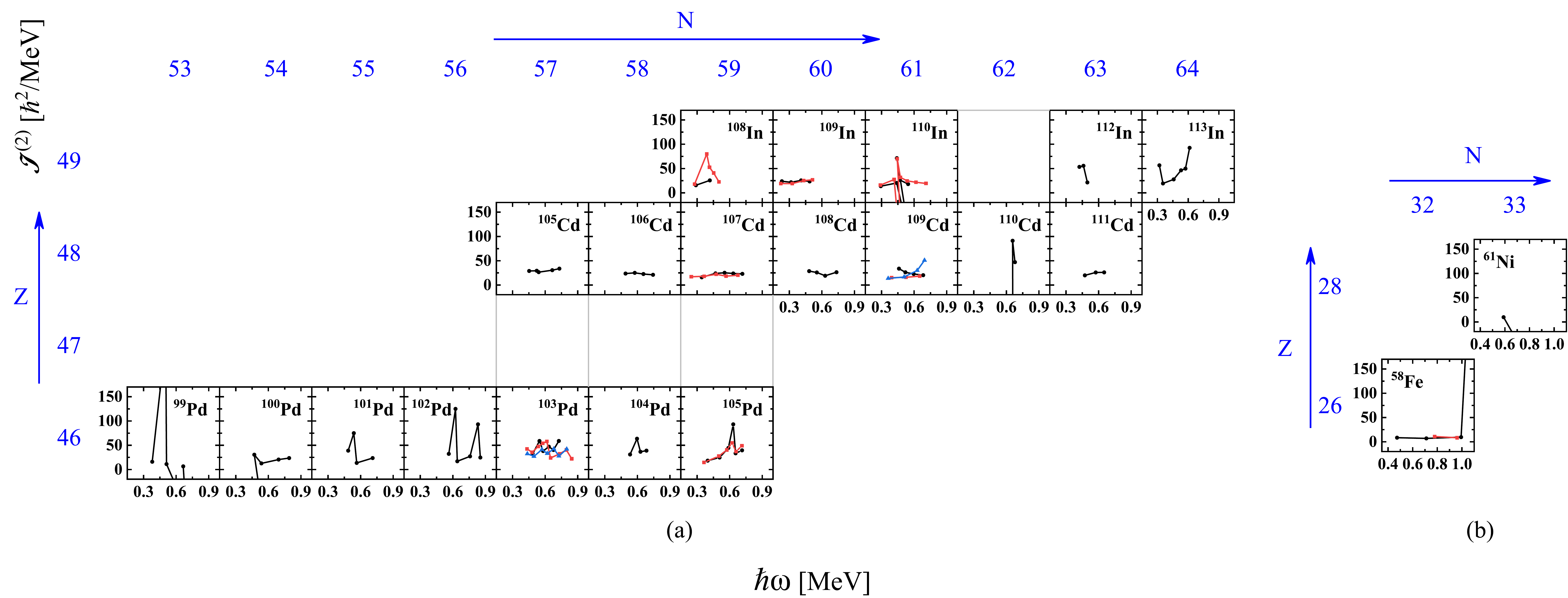}
		\caption{(Color online) Dynamic moment of inertia versus rotational frequency for antimagnetic rotational bands in $A$ $\sim$ (a) 110 and (b) 60 mass regions.}
		\label{31}
	\end{figure}

\begin{figure}[H]
		\centering
		\includegraphics[width=8cm]{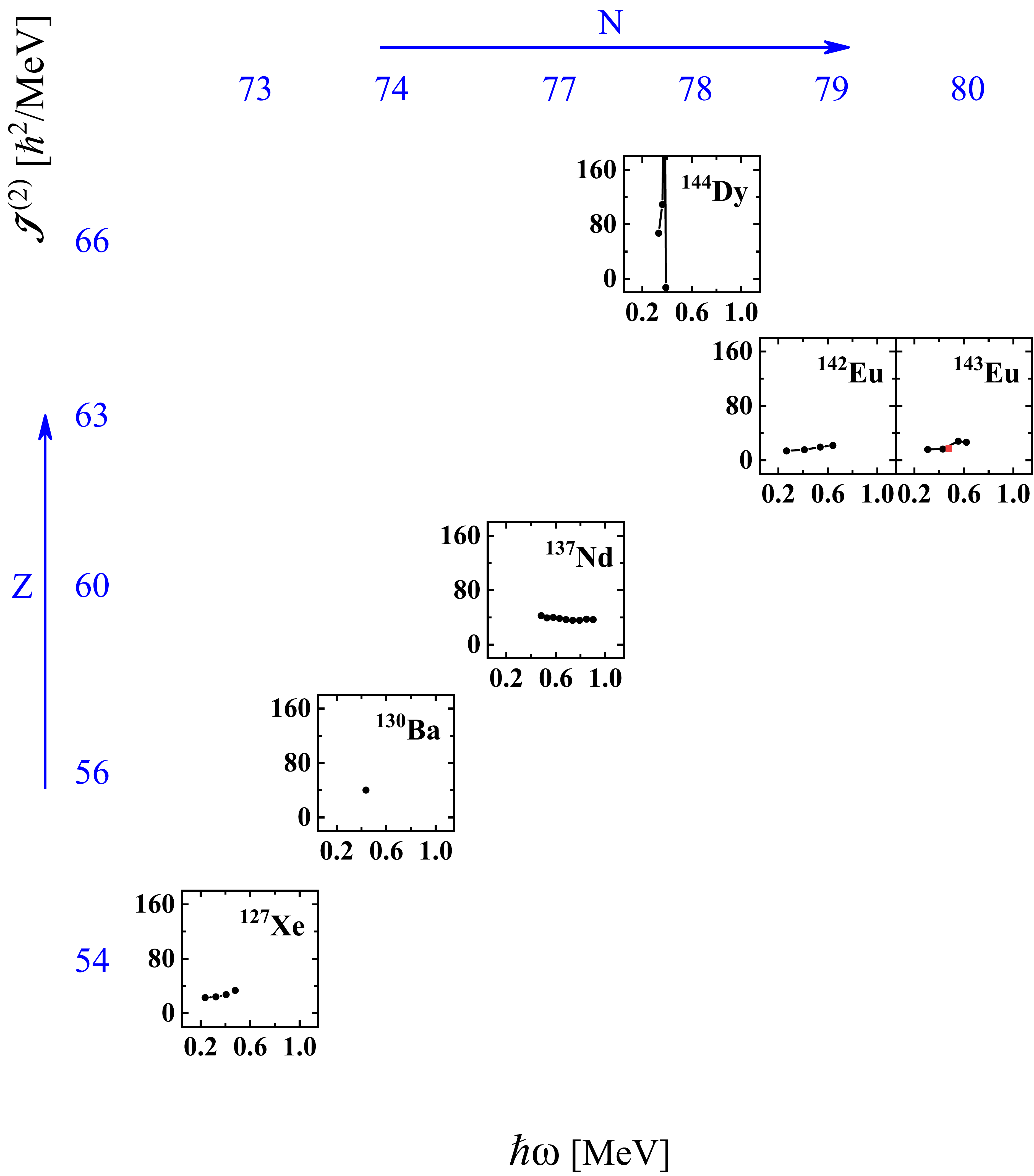}
		\caption{(Color online) Dynamic moment of inertia versus rotational frequency for antimagnetic rotational bands in $A$ $\sim$ 140 mass region.}
		\label{32}
	\end{figure}

	\cref{31,32} show the plots of $\mathcal{J}$$^{(2)}$ vs. $\hbar$$\omega$ for all antimagnetic rotational bands in $A$ $\sim$ 60, 110 and 140 mass regions, respectively. The value of $\mathcal{J}$$^{(2)}$ is estimated by using the relation $\mathcal{J}$$^{(2)}$=$4/[E_\gamma(I+2 \rightarrow I)-E_\gamma(I \rightarrow I-2$)]. The $\mathcal{J}$$^{(2)}$ values are roughly constant with increasing rotational frequency in $^{58}$Fe, $A$ $\sim$ 140 mass region, and Cd isotopes with $A$ $\sim$ 110 mass region. Moreover, there is a small staggering shown in $^{100,~101,~103,~104,~105}$Pd, $^{112}$In, and a large fluctuation is observed in $^{58}$Fe, $^{61}$Ni, $^{99,~102}$Pd, $^{110}$Cd, $^{108,~110,~113}$In, and $^{144}$Dy.
	
	\subsection{The electric quadrupole reduced transition probability vs. spin}
\begin{figure}[H]
		\centering
		\includegraphics[width=15.7cm]{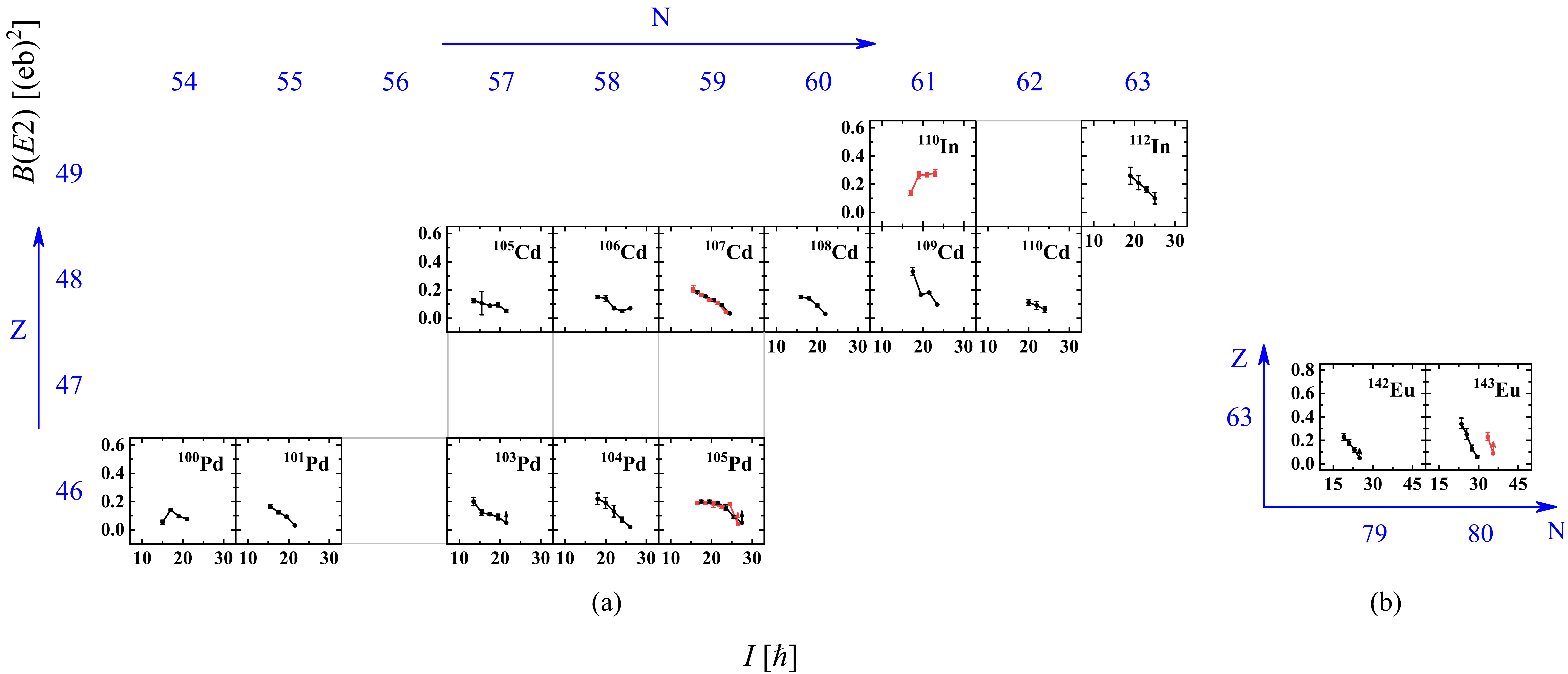}
		\caption{(Color online) Electric quadrupole reduced transition probability versus spin for antimagnetic rotational bands in $A$ $\sim$ (a) 110 and (b) 140 mass regions.}
		\label{35}
	\end{figure}

	The electric quadrupole reduced transition probabilities $B(E2)$ for antimagnetic rotational bands in $A$ $\sim$ 110 and 140 mass regions are shown in \cref{35}. It is obvious that the values of $B(E2)$ are small, less than 0.4 (eb)$^{2}$. In addition, $B(E2)$ values show a decreasing trend with the spin increasing, except for $^{110}$In whose $B(E2)$ values increase as the spin increases. Then, one can note that a small staggering is shown in $^{100}$Pd, $^{105,~106, ~109}$Cd, and $^{110}$In.
	
	\subsection{The $\mathcal{J}$$^{(2)}$/$B(E2)$ ratio vs. spin}
\begin{figure}[H]
		\centering
		\includegraphics[width=15.7cm]{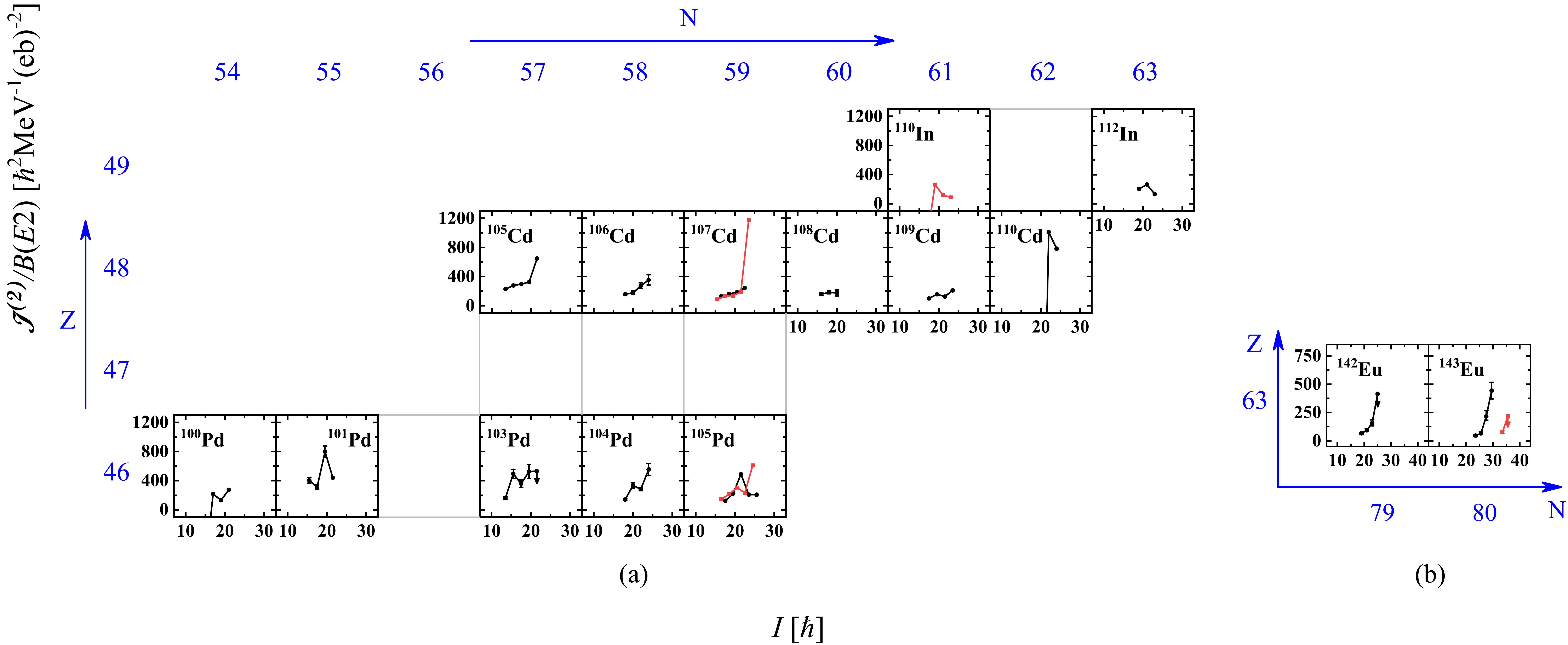}
		\caption{(Color online) $\mathcal{J}$$^{(2)}$/ $B(E2)$ ratio versus spin for antimagnetic rotational bands in $A$ $\sim$ (a) 110 and (b) 140 mass regions.}
		\label{36}
	\end{figure}

	\cref{36} show the plots of $\mathcal{J}$$^{(2)}$/$B(E2)$ vs. $I$ for all antimagnetic rotational bands in $A$ $\sim$ 110 and 140 mass regions. The values of $\mathcal{J}$$^{(2)}$/$B(E2)$ should exceed 100 $\hbar^{2}$MeV$^{-1}$(eb)$^{-2}$ for AMR bands \cite{RN429}. Indeed, the most nuclei satisfy the criterion. However, some $\mathcal{J}$$^{(2)}$/$B(E2)$ values are less than 100 $\hbar^{2}$MeV$^{-1}$(eb)$^{-2}$ in $^{107}$Cd and $^{142,~143}$Eu. Moreover, the $\mathcal{J}$$^{(2)}$/$B(E2)$ roughly grows with the spin increasing except for $^{110}$Cd and $^{110,~112}$In. There is the phenomenon of staggering in Pd isotopes and $^{109}$Cd as well. Due to the limited data in the literatures of antimagnetic rotation, it is difficult to find more details on the relevant properties.
	
	It is well known that AMR bands are expected to be observed in the same regions as MR bands \cite{RN429}. Up to now, AMR bands are observed in $A$ $\sim$ 60, 110, and 140 mass regions, while MR bands are observed not only in the mass regions of $A$ $\sim$ 60, 110, and 140 but also in $A$ $\sim$ 80 and $A$ $\sim$ 190 mass regions. As a result, we can deduce that AMR bands exist in the $A$ $\sim$ 80 and $A$ $\sim$ 190 mass regions as well, which can be investigated further experimentally. In addition, according to the systematics of antimagnetic rotation, we may predict that $^{111}$In could be a candidate nucleus with AMR bands. Then, it is interesting to note that there exist multiple AMR bands in $^{58}$Fe, $^{103,~105}$Pd, $^{107,~109}$Cd, and $^{108,~109,~110}$In.
	\clearpage
	\section{Summary}
	In this paper, 252 magnetic rotational bands reported in 123 nuclei and 38 antimagnetic rotational bands reported in 27 nuclei are collected and listed. Following the presentation of the kinematic moment of inertia $\mathcal{J}$$^{(1)}$, dynamic moment of inertia $\mathcal{J}$$^{(2)}$, and $I$ versus rotational frequency $\omega$, as well as energy staggering parameter $S(I)$, $B(M1)$, $B(E2)$, $B(M1)/B(E2)$, and $\mathcal{J}$$^{(2)}$/$B(E2)$ versus $I$ in $A$ $\sim$ 60, 80, 110, 140, and 190 mass regions, the main features of magnetic and antimagnetic rotational bands are also discussed in detail. 
	
	For magnetic rotation, lots of bands are observed to show the phenomena of band-crossing and signature splitting.
	The kinematic moment of inertia $\mathcal{J}$$^{(1)}$ gradually decreases with the rotational frequency increasing. The regularity of $S(I)$ in $A$ $\sim$ 60, 80, 110, 140, and 190 mass regions is contrary to $\mathcal{J}$$^{(1)}$, which is also well confirmed from the research in this paper. Furthermore, the values of $B(M1)$ are large while the values of $B(E2)$ are small (There is an exception for $^{112}$Sb whose values range between 0.5$-$1.0 (eb)$^{2}$). Both values decrease with increasing spin. The $B(M1)/B(E2)$ values are generally more than 10 [$\mu_{N}$/(eb)]$^{2}$. According to the systematics of magnetic rotation, we may infer that $^{80}$Rb, $^{108}$Ag, $^{114}$Sb, $^{137}$Ce, $^{135,~140}$Nd, $^{140}$Sm, $^{190}$Pb, and $^{196,~201}$Bi could be the candidate nuclei with magnetic rotational bands.

	For antimagnetic rotation, multiple antimagnetic rotational bands in $^{58}$Fe, $^{103,~105}$Pd, $^{107,~109}$Cd, $^{108,~109,~110}$In, and $^{143}$Eu are shown. The spin generally grows steadily as the rotational frequency increases. The kinematic moment of inertia $\mathcal{J}$$^{(1)}$ decreases with the increasing rotational frequency. In addition, the values of $B(E2)$ are small, less than 0.4 (eb)$^{2}$. The $\mathcal{J}$$^{(2)}$/$B(E2)$ roughly grows with the spin increasing, and $\mathcal{J}$$^{(2)}$/$B(E2)$ values should exceed 100 $\hbar^{2}$MeV$^{-1}$(eb)$^{-2}$. The systematic study of antimagnetic rotation suggests that the $^{111}$In may be a candidate nucleus with antimagnetic rotational bands, and $A$ $\sim$ 80 and $A$ $\sim$ 190 mass regions could be predicted as the candidate antimagnetic rotational mass regions. 
	\setcounter{secnumdepth}{0}
	\subsection{Acknowledgments}
	
	The authors are indebted to Prof. G. M. Zeng for the suggestion of this topic and the guidance during this work. This work is supported by the Jilin Scientific and Technological Development Program (20230101009JC), National Natural Science Foundation of China (12175086, 11775098, U1867210, 11405072), Science and Technology Research Planning Project of Jilin Provincial Department of Education (JJKH20220965KJ), Natural Science Foundation of Chongqing, China (CSTB2022NSCQMSX0315), Natural Science Foundation of Sichuan China (23NSFSC1051), and Graduate Innovation Fund of Jilin University (101832020CX080).
	
	\clearpage
	
	\pagestyle{plain}
	\vspace{-3.2em}
	\vspace{0.8em}
	\renewcommand{\baselinestretch}{1.0}
	\setlength{\baselineskip}{18pt}
	
	\appendix
	\bibliographystyle{model1a-num-names}
	\bibliography{1}
	
	
	
	
	



	
	\clearpage
	\setcounter{secnumdepth}{0}
	\subsection{Explanation of Tables}
	\setcounter{secnumdepth}{0}
	\subsubsection{Table 1. Magnetic and antimagnetic rotational bands}
\begin{tabular*}{1\textwidth}{@{}@{\extracolsep{\fill}}lp{6in}@{}}
		\multicolumn{2}{p{0.6\textwidth}}{}\\
		
		$^{A}_{Z}$X$_{N}$	&Denotes the specific nuclide with\\
		& \quad X \quad   Element symbol \\
		&\quad   $A$ \quad   Mass number \\
		&\quad   $Z$   \quad  Proton number \\
		&\quad  $N$   \quad  Neutron number\\
		$E$
		& Level energy in units of keV. Energies in parentheses denote tentative levels. Labels X, Y, Z,
		etc., indicate that absolute excitation energies are unknown due to lack of knowledge about
		linking transitions to the lower levels.\\
		$I$$^{\pi}$
		& $I$ denotes the level spin for each band member. $\pi$ denotes the parity (+ or ${}-$). $I$$^{\pi}$ given in parentheses denotes uncertain spin and/or parity assignments.\\
\end{tabular*}
	\clearpage
	\label{table1}
	\setlength{\tabcolsep}{0.2mm}
	\setlength{\LTcapwidth}{7in}
	\renewcommand{\arraystretch}{0.8}
	\setlength{\columnsep}{1mm}
	
	\setcounter{secnumdepth}{0}
	\subsection{Table A. Magnetic rotational bands. See Explanation of Tables for details.}

	\begin{longtable}{@{\extracolsep\fill}ccccccccccc@{}}
		\hline
		Nuclei &\quad band & $E$ & $I$$^{\pi}$ & $E$$_{\gamma}(M1)$& $E_{\gamma}(E2)$ & $B(M1)$ & $B(E2)$ & $B(M1)/B(E2)$ & References &\\
		&  & keV & $\hbar$ & keV & keV & $\mu_{N}^{2}$ & (eb)$^{2}$ &($\mu_{N}$/eb)$^{2}$ &  &  \\ \hline
		\endfirsthead
		\hline \\
		\endhead
		
		$^{58}$Fe& 1 & 7657.8  & (8)    &     &   &  &  &  & \cite{RN132} &  \\
		&   & 8041.9  & (9)  &  384.1  & & &  &  &  &  \\
		&   & 8540.0  & (10)  & 498.1  & & &  &  &  &  \\
		&   & 9163.6  & (11)  & 623.3  & & &  &  &  &  \\
		&   & 9924.7  & (12)  & 761.2  & & &  &  &  &  \\
		&   & 10952.4 & (13)  & 1027.6 & & &  &  &  &  \\
		&   & 12017.0 & (14)  & 1064.6 & & &  &  &  &  \\
		&   & 13173.4 & (15)  & 1156.7 & & &  &  &  &  \\
\hline

	\end{longtable}
	
	\clearpage 
	
	\label{tableI}
	\setlength{\tabcolsep}{0.6mm}
	\setlength{\LTcapwidth}{7in}
	\renewcommand{\arraystretch}{0.8}
	\setcounter{secnumdepth}{0}
	\subsection{Table B. Antimagnetic rotational bands. See Explanation of Tables for details.}
	
	\begin{longtable}{@{\extracolsep\fill}ccccccccc@{}}
		\hline
		Nuclei & band & $E$ & $I$$^{\pi}$ &  $E_{\gamma}(E2)$ & $B(E2)$ & $\mathcal{J}^{(2)}/B(E2)$ & References &\\
		&  & keV & $\hbar$ & keV  & (eb)$^{2}$ & $\hbar^{2}$MeV$^{-1}$(eb)$^{-2}$ &  &  \\ \hline
		\endfirsthead
		\hline\\
		\endhead
		
		$^{58}$Fe & 1 & 5086.7  & 6     &      &  &  & \cite{RN9058,RN132} &    \\
		&   & 6033.8  & 8     & 947.1  &  &  &  &    \\
		&   & 7457.0  & 10    & 1423.1 &  &  &  &    \\
		&   & 9444.6  & 12    & 1987.5 &  &  &  &    \\
		&   & 11852.7 & (14)  & 2408.4 &  &  &  &    \\
		&   & 14265.0 & (16)  & 2412.0 &  &  &  &    \\
		& 2 & 8424.1  & (10$^{+}$) &      &  &  & \cite{RN9058,RN132} &    \\
		&   & 9983.0  & 12($^{+}$) & 1558.9 &  &  &  &    \\
\hline

	\end{longtable}
	
	\normalsize\bigskip

\end{document}